\documentclass[final,3p,times]{elsarticle}

\usepackage{amssymb}
\usepackage{amsmath}
\usepackage{amsthm}
\usepackage{booktabs}
\usepackage{float}
\usepackage{hyperref}
\hypersetup{hidelinks}
\usepackage{xcolor}
\usepackage{tikz}

\journal{Aerospace Science and Technology}

\newcommand{\dataavailabilitystatement}{%
All benchmark code, per-model configurations, evaluation protocols, and table-generation scripts are openly available at \url{https://github.com/ATMSaxon/U3DWind-Benchmark}. The complete U3DWind dataset at the original 10~m resolution is publicly available through ModelScope at \url{https://modelscope.cn/datasets/ZXX0926/U3DWind}.
}

\newcommand{\publicationnotice}{%
  \par\noindent
  \begingroup
  \setlength{\fboxsep}{5pt}%
  \setlength{\fboxrule}{0.4pt}%
  \fbox{\begin{minipage}{\dimexpr\linewidth-2\fboxsep-2\fboxrule\relax}
    \small\normalfont\raggedright
    \urlstyle{same}%
    This is the Accepted Manuscript version of an article published by Elsevier
    in the journal \textit{Aerospace Science and Technology}. The published
    version is available at:
    \url{https://doi.org/10.1016/j.ast.2026.114092}.
    \par\smallskip
    \textbf{Cite as:} Zhou, S., Wei, H., Xia, C., Xing, Y., Jiang, C., Yang, H.,
    Jia, S. (2027). U3DWind: A low altitude wind field dataset and benchmark for
    urban air mobility. \textit{Aerospace Science and Technology}, 180, 114092.
    \par\smallskip
    \textbf{Data availability:} \dataavailabilitystatement
  \end{minipage}}%
  \endgroup
  \par\vspace{18pt}%
}
\let\originalMaketitleBox\MaketitleBox
\renewcommand{\MaketitleBox}{\publicationnotice\originalMaketitleBox}

\begin{document}

\begin{frontmatter}

\title{{U3DWind}: A low altitude wind field dataset and benchmark for urban air mobility}

\author[1]{Shixiong Zhou\fnref{fn1}}
\author[2]{Huanxia Wei\fnref{fn1}}
\author[3]{Chao Xia}
\author[4]{Yingying Xing}
\author[5]{Changmin Jiang}
\author[6]{Hai Yang}
\author[1,6]{Shuai Jia \corref{cor1} }

\cortext[cor1]{Corresponding author: shuaijia@hkust-gz.edu.cn}
\fntext[fn1]{These authors contributed equally to this work}

%% Author affiliation
\affiliation[1]{organization={Thrust of Intelligent Transportation, The Hong Kong University of Science and Technology (Guangzhou)},
            addressline={1st Duxue Road},
            city={Guangzhou},
            postcode={511543},
            country={China}}
\affiliation[2]{organization={Department of Mechanical and Aerospace Engineering, University of Manchester},
            addressline={Oxford Road},
            city={Manchester},
            postcode={M13 9PL},
            country={UK}}
\affiliation[3]{
		organization={Department of Mechanics and Maritime Sciences},
		addressline={Chalmers University of Technology},
		city={Gothenburg},
		postcode={412 96},
		country={Sweden}}
\affiliation[4]{
  organization={The Key Laboratory of Road and Traffic Engineering of Ministry of Education, Tongji University},
  city={Shanghai},
  postcode={201804},
  country={China}}
\affiliation[5]{organization={Department of Logistics and Maritime Studies, The Hong Kong Polytechnic University},
            addressline={Hung Hom, Kowloon},
            city={Hong Kong},
            country={China}}
\affiliation[6]{organization={Department of Civil and Environmental Engineering, The Hong Kong University of Science and Technology},
            addressline={Clear Water Bay, Kowloon},
            city={Hong Kong},
            postcode={999077},
            country={China}}

%% Abstract
\begin{abstract}
Urban Air Mobility (UAM) requires reliable assessment of low-altitude wind hazards, because winds, gusts, and building-induced turbulence have been recognized as critical factors affecting vehicle stability, route feasibility, vertiport siting, and airspace management. While wind-tunnel experiments, computational fluid dynamics (CFD), multiscale downscaling, reduced-order models, and UAV planning datasets have advanced wind-aware analysis, public resources for data-driven, city-scale UAM planning remain limited in geographic coverage, scenario diversity, vertical extent, building realism, and task-oriented benchmarking. To address this gap, we introduce \textsc{U3DWind}, a building-resolved low-altitude wind-field dataset generated using our GPU-accelerated Lattice Boltzmann Method--Large-Eddy Simulation (LBM-LES) framework for rapid urban flow simulation. \textsc{U3DWind} covers five megacities in China: Beijing, Shanghai, Guangzhou, Shenzhen, and Hong Kong. It contains 720 simulations, with 16 inflow directions, three reference wind speeds, and three seasonal atmospheric scenarios (annual, summer, and winter) for each city. At a 10 m grid resolution, the dataset provides three-dimensional three-component (3D3C) velocity, turbulent kinetic energy (TKE), flow density, and fluid--solid masks. To support operationally relevant evaluation, we further define five baseline tasks: wind-field prediction, sparse-sensor wind-field reconstruction, airworthiness wind-compliance risk scoring, site wind-exposure ranking, and noise propagation modeling. As a multi-city, building-resolved 3D urban wind-field dataset, \textsc{U3DWind} enables systematic evaluation of wind-induced impacts in low-altitude traffic scenarios and provides an open benchmark for urban airspace management and data-driven, building-resolved urban flow modeling.
\end{abstract}

%% Keywords

\end{frontmatter}

\section{Introduction}
\begin{figure}[t]
\centering
\includegraphics[width=1\linewidth]{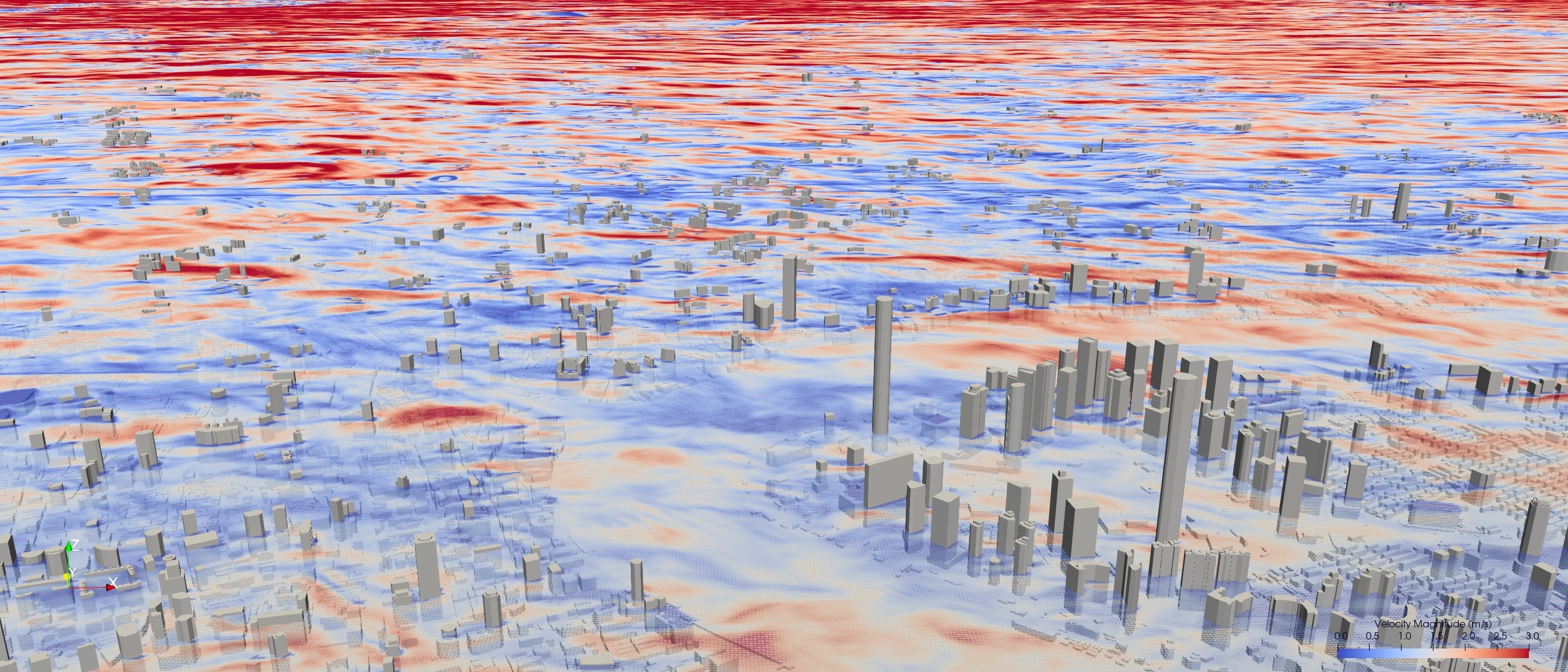}
\caption{Sectional visualization of the instantaneous wind speed over the Shanghai reference domain in the \textsc{U3DWind} dataset. Computed via building-resolved LBM-LES, the rendering captures the heterogeneous urban boundary layer, highlighting aerodynamic interactions such as flow channelization along the Huangpu River corridor and wake generation behind the dense Lujiazui high-rise clusters. This macro-scale complexity underscores the necessity of 3D building-resolved aerodynamics for low-altitude UAM operations.}
\label{fig:hero_shanghai}
\end{figure}

With the rapid advancement of Urban Air Mobility (UAM), an increasing number of cities have begun to actively plan and design their future urban airspace \citep{Bauranov2021UAMAirspace,Johnson2022NASAConcept}. Recent assessments further frame UAM as an integrated low-altitude airspace system that requires coordinated planning of routes, vertiports, and operational constraints \citep{PonsPrats2022UAMStatus}. Following the developmental trajectory of traditional civil aviation, safe UAM integration requires early evaluation of environmental and operational hazards \citep{Mueller2017UAMAirspace,Reiche2021UAMWeather}. Low-altitude wind is a central component of this evaluation. Wind-tunnel experiments with scaled urban models have shown that flow acceleration, directional variability, shear, and turbulence around and between buildings can challenge UAS aerodynamic stability \citep{AlLabbad2022UrbanAirflowUAM}. Simulation-based AAM studies further indicate that urban wind fields can affect control effort and vehicle response, especially for lower wing-loading aircraft \citep{Krawczyk2024ReducedOrderAAM}. In dense urban airspace, these effects are amplified by invisible turbulence induced by diverse building morphologies (Fig. \ref{fig:hero_shanghai}), which can threaten low-altitude flight safety \citep{Reiche2021UAMWeather}. Furthermore, while noise pollution is generally confined to airport perimeters in traditional aviation, the high-frequency, low-altitude nature of UAM operations brings these acoustic impacts directly into densely populated communities \citep{Cetin2022UAMAcceptance,Pascioni2018UAVNoise}. Consequently, UAM planning requires wind and noise representations at spatial and vertical resolutions sufficient for route safety, vertiport siting, and operational impact assessment.

Researchers have therefore developed low-altitude urban wind modeling methods across several traditions. Wind engineering studies use CFD and wind tunnel testing to resolve building-scale flow features \citep{Blocken2016Review,Mochida2008Pedestrian}. Practical CFD guidelines document the sensitivity of such predictions to domain size, boundary conditions, grid arrangement, and turbulence closure \citep{Tominaga2008AIJGuidelines}. Urban microclimate reviews show that CFD has become a common predictive tool for building-resolved outdoor environments \citep{Toparlar2017CFDUrbanMicroclimate}. Downscaling work maps mesoscale wind conditions into the urban canopy using microscale models \citep{Ricci2022StaticDownscaling}. UAM-oriented studies have also begun to use reduced-order realistic urban winds to evaluate aircraft response \citep{Vuppala2024AAMUrbanWinds}, energy-aware UAV planning with LES-generated urban wind fields \citep{Rienecker2023EnergyOptimalUAV}, and Gappy POD reconstruction from sparse wind sensors \citep{Ebert2023GappyPODWindPlanning}. These efforts establish a methodological foundation for wind-aware UAM analysis, while exposing a common limitation in the data layer. Most publicly accessible urban wind datasets remain tailored to micro-scale applications, such as pedestrian-level wind comfort or single-building aerodynamics \citep{Blocken2012Pedestrian}. Existing UAM-oriented data resources typically cover local domains, reduced-dimensional fields, or limited inflow scenarios \citep{Baskar2020WindDataset}. Meanwhile, large-scale, high-resolution city simulations with conventional CFD solvers remain computationally expensive \citep{Maronga2020PALM}. This combination of limited public data and high simulation cost constrains the development of data-driven methods for UAM operations.

Motivated by this modeling and data need, we introduce \textsc{U3DWind}, an extensive and publicly accessible three-dimensional urban wind dataset tailored specifically for Urban Air Mobility research. Generated via our \textsc{LatticeUrbanWind} (LUW) framework \citep{LUW}, this dataset encompasses the five megacities: Beijing, Shanghai, Guangzhou, Shenzhen and Hong Kong. To anchor the simulations in realistic operating conditions, the cases couple NASA POWER-derived meteorological inflows with building footprints jointly extracted from OpenStreetMap \citep{Boeing2017OSMnx} and the Microsoft GlobalMLBuildingFootprints dataset, and terrain elevations from the Copernicus GLO-30 Digital Elevation Model. Furthermore, to demonstrate the practical utility of the dataset, we establish a comprehensive benchmark based on the Shanghai subset. This benchmark features five fundamental urban computing applications: wind field prediction, wind field reconstruction using sparse sensors, airworthiness wind-compliance risk scoring, site wind-exposure ranking, and noise propagation modeling.
The primary contribution of this work is a curated simulation dataset together with a standardized benchmark protocol: the flow fields are generated with our previously released \textsc{LatticeUrbanWind} solver \citep{LUW}, and the five tasks are designed as controlled test beds for model evaluation and method development. Our contributions are threefold. First, we introduce \textsc{U3DWind}, an open-source, decameter-resolution (10 m) three-dimensional wind field dataset designed for urban air mobility (UAM). Covering five representative megacities in China, \textsc{U3DWind} extends beyond the single-city or idealized settings commonly considered in existing datasets and helps bridge the scale gap between mesoscale weather forecasting and microscale urban aerodynamics. It therefore provides high-resolution, low-altitude wind information that is directly relevant to flight operations in complex urban environments.
Second, we establish a standardized benchmark to evaluate the practical utility of \textsc{U3DWind} under operationally meaningful UAM scenarios. This benchmark defines five baseline tasks, including 3D wind field prediction, wind field reconstruction from sparse sensors, airworthiness wind-compliance risk scoring, site wind-exposure ranking, and UAV noise propagation modeling. Tasks 1 and 2 (wind-field prediction and sparse-sensor reconstruction) form the core wind-field benchmark that is directly defined by the released fields; Tasks 3, 4, and 5 are downstream application illustrations that introduce explicitly stated operational assumptions on airworthiness thresholds, siting criteria, and acoustic source models. Together with unified evaluation protocols, these tasks provide a systematic basis for assessing model performance across safety-critical and infrastructure-oriented applications.
Third, \textsc{U3DWind} enables quantitative assessment of future low-altitude transportation infrastructure at a cross-city scale. By providing unified evaluation metrics and a comprehensive suite of baseline models, it lowers the entry barrier for researchers without specialized backgrounds in fluid dynamics and allows experts in computer science, transportation, and urban systems to investigate flight safety, vertiport siting, and community noise impacts. In this way, the dataset and benchmark jointly support the data-driven planning and evaluation of future UAM networks.

The remainder of this paper is organized as follows. Section \ref{relatedwork} reviews the related work. Section \ref{benchmark} details the construction of the dataset and defines the benchmark tasks. Section \ref{sec:4} presents the experimental setup and evaluation results. Section \ref{sec:5} discusses the broader implications of this work, along with its limitations and future directions. Finally, our key conclusions are drawn in Section \ref{sec:6}.

\section{Related Work}
\label{relatedwork}

Wind-aware UAM research has expanded rapidly from weather-barrier assessment to vehicle-response modeling and operational decision support. Reviews of UAM wind-flow modeling synthesize the need for microscale urban flow representations \citep{Nithya2024UAMWindReview}. Experimental work with scaled urban models has measured wind speed changes, direction shifts, shear, and turbulence intensity that are relevant to UAS operational limits \citep{AlLabbad2022UrbanAirflowUAM}. Flight-dynamics studies driven by realistic urban wind fields show that reduced-order models can reproduce some trajectory-level responses, while detailed spatiotemporal wind content remains important for control-use statistics and saturation events \citep{Krawczyk2024ReducedOrderAAM}. These developments make data availability a central issue for the next stage of UAM wind modeling. A recent review consolidates the open issues of urban wind environments for safe UAM utilization \citep{Jia2026TenQuestions}, and wind-tunnel studies of rooftop wind with parapets quantify the sheltering effects that shape rooftop landing-site exposure \citep{Lin2025Parapet,Lin2026ParapetSheltering}.

At the methodological level, the simulation of urban aerodynamics traditionally relies on computational fluid dynamics, with model forms ranging from Reynolds Averaged Navier Stokes (RANS) equations \citep{Xie2006LESRANS} to large eddy simulations (LES) that explicitly resolve individual buildings \citep{Maronga2020PALM}. Highly accurate simulation frameworks have become reference standards for capturing complex turbulence within the urban boundary layer \citep{Blocken2016Review}. Urban atmospheric-flow reviews further emphasize the interaction among synoptic forcing, terrain, building geometry, and canopy-scale turbulence \citep{Fernando2010UrbanAtmospheres}. High-resolution urban LES studies demonstrate that street-canyon and neighborhood-scale simulations can resolve flow structures that simplified models tend to miss \citep{Letzel2008UrbanLES}. However, these methodologies face a structural contradiction between simulation fidelity and the demand for immense data volume. Running a single simulation at the dimension of an entire city with a resolution of mere meters on conventional unstructured solvers incurs tens to hundreds of thousands of computing hours. Consequently, existing computational campaigns are persistently restricted to isolated neighborhoods \citep{Mochida2008Pedestrian} or a handful of inflow scenarios \citep{Blocken2012Pedestrian}. This severe computational bottleneck means that traditional outputs are rarely assembled into the vast, heterogeneous, and richly annotated datasets required to train and evaluate modern aerodynamic models fueled by machine learning.

The lattice Boltzmann method offers a fundamentally different kinetic approach to fluid dynamics, where local stream and collide operations map exceptionally well to the parallel architecture of modern graphics processing units \citep{King2017GPULBMUrban,Lenz2019LBMRealtime}. Recent advancements in implementations optimized for these processors have enabled simulations comprising a billion cells at interactive execution rates \citep{Jacob2018LBMComfort,Merlier2019LBMPollutant}. Despite this leap in computational throughput, a critical gap remains in the data ecosystem. Most existing urban studies using this method focus narrowly on algorithm validation or isolated dispersion events \citep{Ahmad2017GustLBM}. The community has yet to fully exploit this approach as an engine for massive data generation. Prior to this work, no openly released dataset spanning multiple cities provided the volumetric velocity fields at decameter resolution necessary for urban air mobility.

Public wind-field data also remain misaligned with UAM requirements. Aerial vehicles operate in the lowest few hundred meters of the urban airspace, where wind fields are dictated by a combination of synoptic weather patterns and highly irregular building topologies \citep{Reiche2021UAMWeather}. To simulate this environment, researchers commonly drive microscale solvers with mesoscale numerical weather prediction outputs or global reanalysis products \citep{Olauson2018ERA5}. Building-resolved microscale frameworks such as PALM have substantially improved the physical representation of urban canopy flows \citep{Maronga2020PALM}. Yet many public datasets rely on idealized building arrays that omit true topography \citep{Xie2006LESRANS}, derive from meteorological observations at spatial resolutions far coarser than actual buildings, or cover local two-dimensional path-planning settings \citep{Baskar2020WindDataset}. Such simplifications reduce computational cost, while also filtering out morphology-induced turbulence such as street-canyon jets and rooftop separation. A dataset capable of supporting operational safety evaluations must therefore be anchored in authentic urban topologies \citep{Boeing2017OSMnx} and real weather climatologies.

The integration of deep learning with fluid mechanics is accelerating rapidly, utilizing convolutional networks, Fourier neural operators \citep{Li2021FNO,Wen2022UFNO}, and models guided by physics \citep{Raissi2019PINN,Cai2021PINNFluid} to emulate fluid dynamics. Operator-learning frameworks such as DeepONet \citep{Lu2021DeepONet} and global weather emulators such as FourCastNet \citep{Pathak2022FourCastNet} have further demonstrated the capacity of these architectures for high-dimensional fluid systems. However, algorithmic evolution has significantly outpaced the supply of robust data. Comprehensive surveys consistently identify data scarcity and a lack of environmental diversity as dominant bottlenecks in data-driven flow modeling \citep{Cai2021PINNFluid} and UAM-oriented wind analysis \citep{Nithya2024UAMWindReview}. At present, most predictive models are trained on fragmented, internal datasets specific to a single city or an idealized layout. This pervasive isolation of data prevents fair methodological comparisons and obscures whether these models can genuinely generalize to new urban environments. To transition aerodynamic risk assessment from isolated experiments to reliable infrastructure planning, a benchmark integrating multiple cities, authentic geometries, and complete volumetric wind velocity fields is critically required~\citep{Zhou2026RooftopRecon}.

\section{\textsc{U3DWind} Benchmark}
\label{benchmark}

This section details the methodology for constructing the \textsc{U3DWind} dataset, outlines its physical specifications, and mathematically formalizes the five benchmark tasks designed to evaluate data-driven applications in UAM.

\subsection{Data Generation Pipeline}
\label{subsec:pipeline}

The \textsc{U3DWind} dataset is generated using \textsc{LatticeUrbanWind}, our open-source GPU-accelerated lattice Boltzmann large-eddy simulation (LBM-LES) framework \citep{LUW}. To accurately capture urban aerodynamics, the simulation pipeline integrates real geographical and meteorological data. Urban topologies, including building footprints and heights, are extracted jointly from OpenStreetMap (OSM) \citep{Boeing2017OSMnx} and the Microsoft GlobalMLBuildingFootprints dataset, while terrain elevations are derived from the Copernicus GLO-30 Digital Elevation Model (DEM). These spatial inputs are subsequently reprojected and voxelized onto a uniform Cartesian grid with a 10 m resolution. These geometry sources carry their own uncertainties, missing or outdated footprints and generalized building heights in OpenStreetMap and the Microsoft footprints, and terrain smoothing in the 30~m DEM, which propagate into the voxelized geometry and are inherited by the simulated fields. Furthermore, the inflow boundary conditions are driven by a decade of NASA POWER meteorological reanalysis data. Representative power-law wind speed profiles and local wind roses are derived from this historical climatology, ensuring that the simulated cases are tied to observed urban wind climatology and realistic boundary-layer structure \citep{Ahmad2017GustLBM}. For lattice verification as well as validations against observations, see Supplementary Material S1.

\subsubsection{Numerical setup and solver configuration}
\label{subsubsec:numsetup}

The \textsc{LatticeUrbanWind} core is a multi-GPU lattice Boltzmann solver built on
FluidX3D: a D3Q19 lattice with a single-relaxation-time (SRT) collision operator,
distribution functions stored in the compressed FP16C format with all arithmetic in
FP32, and a Smagorinsky--Lilly subgrid-scale model for the unresolved turbulence.
Inflow and outflow faces use equilibrium boundaries; solid cells voxelized from the
building and terrain geometry impose no-slip conditions through halfway bounce-back,
without wall functions. The temperature extension is enabled to impose the seasonal
atmospheric scenarios, and the Coriolis body force is active in all runs. Terrain is
voxelized from the Copernicus GLO-30 elevation model by inverse-distance weighting
over the 12 nearest samples, and building footprints are extruded from the merged
OpenStreetMap and Microsoft GlobalMLBuildingFootprints polygons.

Each case runs for 50{,}000 time steps on a fixed $10$~m lattice; the first 2{,}000
steps are discarded as spin-up and the released mean fields and TKE accumulate over
the remaining window with a 10-step sampling stride. Synthetic turbulent inflow is
generated by a von K\'arm\'an spectral method (turbulence intensity $5\%$, integral
length scale $100$~m, 256 modes, vertical anisotropy factor 5). Lateral buffer
nudging (160~m, $\tau = 300$~s) and a top sponge layer (200~m, $\tau = 120$~s)
suppress recirculation artifacts at the domain boundaries. Table~\ref{tab:cfd_setup}
summarizes the shared numerical configuration.

At this resolution the released fields resolve building-scale wakes, channeling along
river and street corridors, and canopy-layer shear at scales of two to three cells
($20$--$30$~m) and larger. They do not resolve the near-wall viscous sublayer,
separation features below the $10$~m cell size, street furniture, or
thermally driven microcirculations beyond the scenario-level temperature forcing;
the absence of wall functions is a known model-form uncertainty that is inherited by
every downstream benchmark number.

\begin{table}[htbp]
\centering
\caption{Numerical configuration shared by all 720 \textsc{U3DWind} runs; per-city
lateral domain extents are listed in Table~\ref{tab:cross-city}.}
\label{tab:cfd_setup}
\footnotesize
\begin{tabular}{lc}
\toprule
Domain extents & lateral per city as in Table~\ref{tab:cross-city}; $L_z = 1.1$~km for all cities \\
Lattice / collision & D3Q19, SRT, FP16C storage (FP32 arithmetic) \\
Subgrid model & Smagorinsky--Lilly LES \\
Walls & no-slip halfway bounce-back, no wall functions \\
Inflow / outflow & equilibrium boundaries, von K\'arm\'an synthetic inflow \\
Grid resolution & $10$~m, uniform Cartesian \\
Run length & 50{,}000 steps; spin-up discard 2{,}000; sampling stride 10 \\
Boundary treatment & lateral nudging 160~m ($\tau=300$~s); top sponge 200~m ($\tau=120$~s) \\
Physics & Coriolis force; temperature extension for seasonal scenarios \\
GPU decomposition & $[4,2,1]$, eight GPUs per run \\
\bottomrule
\end{tabular}
\end{table}

\subsubsection{Wind-tunnel validation and grid convergence}
\label{subsubsec:aijval}

The solver configuration is validated against
the Architectural Institute of Japan (AIJ) Case A benchmark of turbulent flow around an
isolated 1:1:2 square prism (Meng--Hibi experiment; building width $B=32$~m,
height $H=64$~m at full scale, roof-height reference speed
$U_H = 4.49$~m/s), the standard validation case for building-resolved
urban CFD \citep{Tominaga2008AIJGuidelines,MengHibi1998}. The case is run at
three lattice resolutions, 2, 4, and 10~m ($H/\Delta = 32$, $16$, and $6.4$),
the coarsest matching the \textsc{U3DWind} production grid, and compared with
the wind-tunnel measurements on two horizontal planes ($z/H = 0.125$ and
$1.25$) and the centreline vertical section. Table~\ref{tab:aij} reports the
fraction-of-two agreement (FAC2), the hit rate $q$ at tolerance $0.25$, the
correlation $r$, and the normalized root mean squared error (NRMSE) for the normalized mean speed and TKE.

\begin{table}[htbp]
\centering
\caption{AIJ single-prism validation at three lattice resolutions. $q$ is the
hit rate at tolerance $0.25$. The TKE FAC2 excludes near-zero measurement
points ($|k/U_H^2| \le 0.05$), leaving 26 to 30 of the 60 or 66 points per
plane; the hit rate is not reported for TKE because the $0.25$ tolerance is
not discriminative for normalized TKE magnitudes.}
\label{tab:aij}
\footnotesize
\begin{tabular}{llcccccc}
\toprule
 & & \multicolumn{4}{c}{Mean speed $U/U_H$} & \multicolumn{2}{c}{TKE $k/U_H^2$} \\
\cmidrule(lr){3-6}\cmidrule(lr){7-8}
Grid & Plane & FAC2 & $q$ & $r$ & NRMSE & FAC2 & $r$ \\
\midrule
2~m  & $z/H=0.125$ & 0.74 & 0.62 & 0.78 & 0.26 & 0.31 & 0.64 \\
2~m  & $z/H=1.25$  & 0.93 & 0.97 & 0.92 & 0.13 & 0.90 & 0.88 \\
2~m  & vertical    & 0.81 & 0.89 & 0.97 & 0.12 & 0.90 & 0.53 \\
\midrule
4~m  & $z/H=0.125$ & 0.12 & 0.28 & 0.67 & 0.43 & 0.00 & 0.25 \\
4~m  & $z/H=1.25$  & 0.83 & 0.87 & 0.87 & 0.16 & 0.45 & 0.68 \\
4~m  & vertical    & 0.70 & 0.76 & 0.94 & 0.15 & 0.50 & 0.13 \\
\midrule
10~m & $z/H=0.125$ & 0.03 & 0.25 & 0.79 & 0.52 & 0.00 & $-0.51$ \\
10~m & $z/H=1.25$  & 0.81 & 0.80 & 0.72 & 0.21 & 0.00 & $-0.39$ \\
10~m & vertical    & 0.60 & 0.67 & 0.88 & 0.18 & 0.00 & $-0.42$ \\
\bottomrule
\end{tabular}
\end{table}

At 2~m the solver reproduces the measurements across all planes (above-roof
FAC2 $0.93$, hit rate $0.97$, vertical-section correlation $0.97$),
establishing the correctness of the numerical scheme. At the 10~m production
resolution the above-roof layer and the vertical section still meet the common
hit-rate acceptance level of $q \geq 0.66$
\citep{Tominaga2008AIJGuidelines}, while the near-ground plane at
$z/H = 0.125$ (8~m above ground, below one lattice cell) is not resolved, and
the measured TKE field around the isolated tower is not reproduced
(FAC2 $= 0$, negative correlation).

These results bound what the released fields can support: mean-wind exposure
at UAM-relevant altitudes rests on the resolved regime, whereas street-level
flow below roughly one cell height and point-accurate TKE around individual
towers do not, so the released TKE is a resolved-scale turbulence indicator; the Task 3 gust proxy inherits this caveat
(Section~\ref{subsec:limits}). Extended profiles and the lattice verification
are provided in Supplementary Material S1.

\subsection{Dataset Specifications}
\label{subsec:specifications}

\begin{figure}[htbp]
    \centering
    \includegraphics[width=0.9\linewidth]{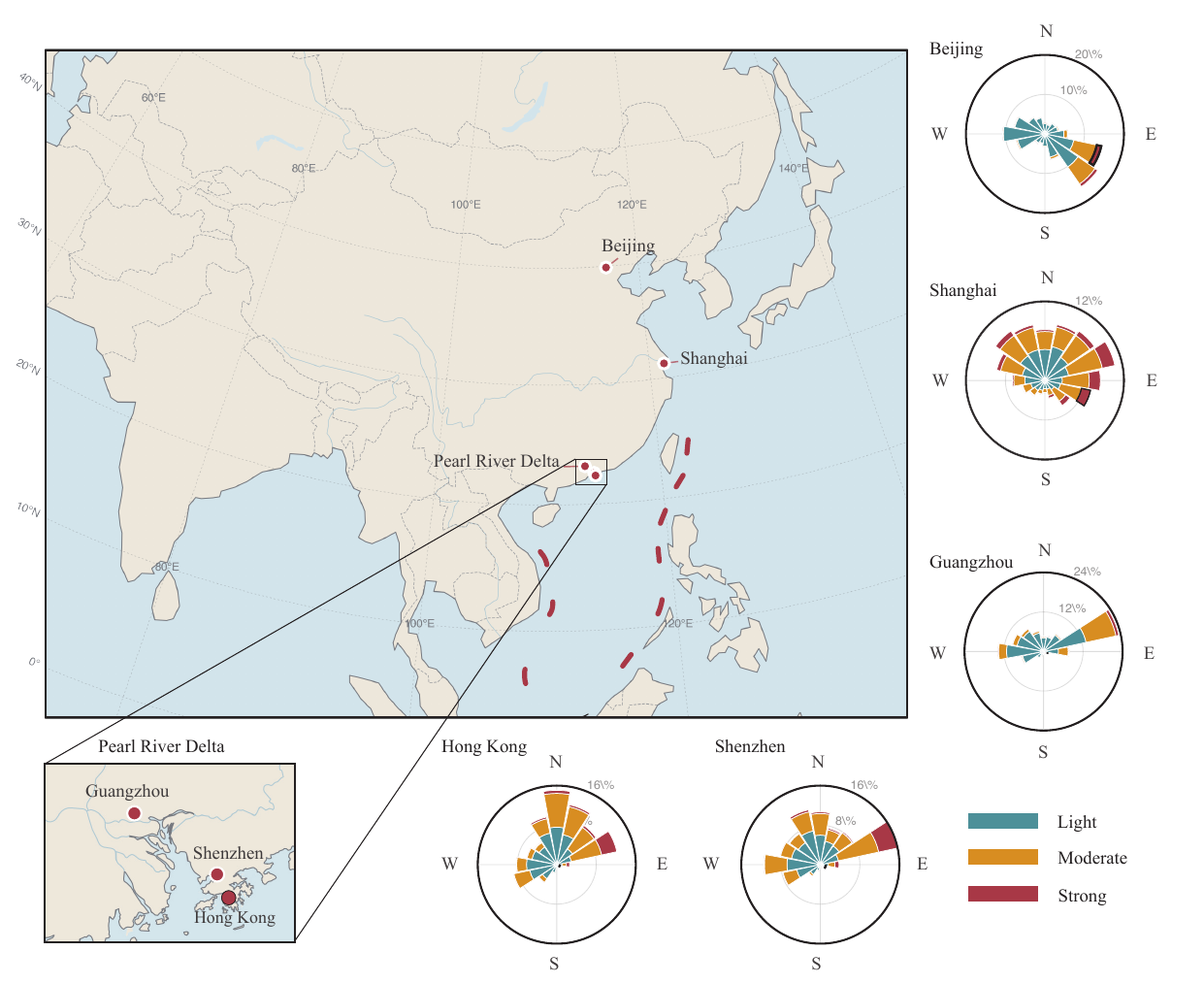}
    \caption{Geographic footprint of the \textsc{U3DWind} release and per-city annual wind climatology. The base map (Lambert-conformal projection) places the five \textsc{U3DWind} cities on mainland China and Hong Kong; each city marker is paired with a 16-sector annual wind rose drawn at the corresponding inset, with sector mass color-coded by wind-speed tier (calm, light, moderate, strong). The annual rose is the simple mean of the SUMMER (Apr--Sep) and WINTER (Oct--Mar) seasonal climatologies derived from NASA POWER 2014--2024 daily reanalysis at the city centroid.}
    \label{fig:cities}
\end{figure}

The dataset encompasses the central domains of five Chinese megacities: Beijing, Shanghai, Guangzhou, Shenzhen, and Hong Kong. For each urban domain, parametric simulations are conducted across 16 inflow azimuths, 3 reference wind speeds ($3$, $6$, and $9~\mathrm{m/s}$), and 3 atmospheric scenarios (annual mean, summer, and winter), yielding a total of 720 stationary flow cases. Table~\ref{tab:dataset_spec} summarizes the per-city grids, case counts, and released variables. The geographical distribution of these five cities, along with their respective wind roses, are illustrated in Fig. \ref{fig:cities}. The release provides the time-averaged three-dimensional, three-component (3D3C) velocity field $\bar{\mathbf{u}}$, turbulent kinetic energy $k$, fluid density $\rho$, and a binary geometric mask $\chi$.

The five released cities span meaningfully different urban morphologies and inflow climatologies, so that downstream studies can probe the cross-city generalization of data-driven urban wind models. Table~\ref{tab:cross-city} summarizes the dominant morphological and climatological descriptors for each city. Domain extents are city specific: Beijing $50 \times 40$\,km, Shanghai $45 \times 35$\,km, Guangzhou $40 \times 35$\,km, Shenzhen $35 \times 25$\,km, and Hong Kong $30 \times 25$\,km.

\begin{table}[htbp]
\centering
\caption{Morphological and inflow-climatological descriptors of the five \textsc{U3DWind} cities within the released computational domain. Building statistics are evaluated on the voxelised solid mask within each city's specific domain; wind-climate descriptors are evaluated on the seasonal wind roses derived from NASA POWER 2014 to 2024 daily reanalysis.}
\label{tab:cross-city}
\footnotesize
\begin{tabular}{lccccc}
\toprule
Descriptor & Beijing & Shanghai & Guangzhou & Shenzhen & Hong Kong \\
\midrule
Domain extent $L_x \times L_y$ (km) & $50 \times 40$ & $45 \times 35$ & $40 \times 35$ & $35 \times 25$ & $30 \times 25$ \\
Terrain relief $\Delta z_{95-5}$ (m) & 92 & 12 & 74 & 188 & 410 \\
Building plan-area fraction (\%) & 24.8 & 33.6 & 28.2 & 31.4 & 22.5 \\
Median building height $H_{50}$ (m) & 24 & 30 & 26 & 36 & 38 \\
95th-percentile building height $H_{95}$ (m) & 112 & 188 & 146 & 218 & 232 \\
High-rise share ($H > 100$\,m, \%) & 2.1 & 6.8 & 3.9 & 7.6 & 8.4 \\

Dominant summer sector & S & SE & S & S & SSW \\
Dominant winter sector & NW & NNE & NNE & NNE & NNE \\
Power-law exponent $\alpha$ (annual) & 0.242 & 0.171 & 0.241 & 0.140 & 0.143 \\
Power-law exponent $\alpha$ (summer) & 0.230 & 0.166 & 0.257 & 0.149 & 0.154 \\
Power-law exponent $\alpha$ (winter) & 0.254 & 0.177 & 0.225 & 0.131 & 0.131 \\
\bottomrule
\end{tabular}
\end{table}

\begin{table}[htbp]
\centering
\caption{\textsc{U3DWind} release specification. Grid dimensions are the 10~m
lattice $(n_x \times n_y \times n_z)$; every city carries 144 cases (16
directions $\times$ 3 speeds $\times$ 3 scenarios) with time-averaged 3D3C
velocity, TKE, density, and the fluid--solid mask. The complete dataset
at the original 10~m resolution is publicly available through ModelScope,
as detailed in the Data availability statement.}
\label{tab:dataset_spec}
\footnotesize
\begin{tabular}{lcc}
\toprule
City & 10~m grid & Cases \\
\midrule
Beijing   & $5000 \times 4000 \times 110$ & 144 \\
Shanghai  & $4500 \times 3500 \times 110$ & 144 \\
Guangzhou & $4000 \times 3500 \times 110$ & 144 \\
Shenzhen  & $3500 \times 2500 \times 110$ & 144 \\
Hong Kong & $3000 \times 2500 \times 110$ & 144 \\
\bottomrule
\end{tabular}
\end{table}

Together, these five cities encompass a diverse range of geographic topologies: Beijing represents a basin-bounded northern plain; Shanghai, a flat coastal delta; Guangzhou, a subtropical river delta; Shenzhen, a mountain-coastal transition zone along the Pearl River Estuary; and Hong Kong, a bay-island archipelago. The corresponding flow climatologies span continental winter monsoon reversal (Beijing), maritime monsoon over an estuary mouth (Shanghai), monsoonal South China coast (Guangzhou), zonal land-sea breeze interaction (Shenzhen), and channeled bay flow over rugged terrain (Hong Kong). These contrasts produce distinct building-wake regimes and distinct prevailing flow channels across the five domains.

The inflow boundary condition for each LBM-LES case is a 1D power-law wind profile $U(z) = u_{10}\,(z/10)^{\alpha}$ driven by NASA POWER 2014--2024 daily reanalysis at the city center. For each city, the per-city power-law exponent $\alpha$ in Table~\ref{tab:cross-city} is fitted from the climatological mean of the 10\,m and 50\,m wind speeds (one mean per season; the annual value is the average of the WINTER and SUMMER fits). The reference 10\,m wind speed ($u_{10}$) is evaluated at three operationally meaningful levels: 3, 6, and 9\,m/s. These values bracket the typical operating envelope of small electric vertical take-off and landing (eVTOL) aircraft, representing a routine breeze, a moderate operational wind, and a near no-fly threshold, respectively. For each (city, $u_{10}$, scenario) triple the same power-law shape is rotated through the 16 azimuth bins, while a von K\'arm\'an synthetic inlet generates the resolved turbulence above the rooftop layer. Three atmospheric scenarios are released per (city, $u_{10}$): an annual scenario using the average of the WINTER and SUMMER $\alpha$, a summer scenario using the SUMMER $\alpha$, and a winter scenario using the WINTER $\alpha$ (Table~\ref{tab:cross-city}). The 16 directions $\times$ 3 speed tiers $\times$ 3 scenarios $\times$ 5 cities therefore yield the 720 stationary cases that compose the \textsc{U3DWind} dataset.

\subsection{Benchmark Tasks Definition}
\label{subsec:formulation}
To standardize the evaluation of data-driven UAM applications, we mathematically formalize five benchmark tasks over the 3D urban computational domain $\Omega \subset \mathbb{R}^3$. The operational objectives, explicit input/output tensor definitions, and rigorous mathematical formulations for all downstream tasks are comprehensively synthesized in Table \ref{tab:benchmark_tasks}. Tasks 1 and 2 operate directly on the released wind fields and constitute the core benchmark; Tasks 3, 4, and 5 are downstream illustrations whose additional assumptions are stated in their respective subsections and whose labels remain fully derivable from the released data.

\begin{table}[htbp]
\centering
\caption{Comprehensive formulation of the \textsc{U3DWind} benchmark tasks. The table details the operational objectives, tensor boundaries, and mathematical mappings for each downstream UAM application.}
\label{tab:benchmark_tasks}
\footnotesize
\setlength{\tabcolsep}{4pt}
\begin{tabular}{p{0.165\textwidth} p{0.23\textwidth} p{0.275\textwidth} p{0.25\textwidth}}
\toprule
\textbf{Benchmark Task} & \textbf{Operational Objective} & \textbf{Inputs \& Outputs} & \textbf{Mathematical Formulation} \\ \midrule

\textbf{1. Surrogate \newline Modeling} &
Emulate LBM-LES to achieve instantaneous 3D aerodynamic field generation. &
\textbf{In:} Geometric mask $g \in \{0,1\}^{N_x \times N_y \times N_z}$; Inflow conditions $c = (U_{\infty}, \theta_{\infty})$ \newline
\textbf{Out:} 3D3C velocity $\hat{\mathbf{u}}(\mathbf{x}) \in \mathbb{R}^3$; Turbulent kinetic energy $\hat{k}(\mathbf{x}) \in \mathbb{R}$ &
$(\hat{\mathbf{u}}, \hat{k}) = f_{\theta}(g, c)$ \\ \midrule

\textbf{2. Sparse \newline Reconstruction} &
Recover global dense flow fields from a deterministic operational sensor pool spanning rooftops and street-level corridor anchors. &
\textbf{In:} Sparse velocity observations $\mathbf{U}_s$; sensor spatial coordinates $\mathbf{X}_s$ drawn from the fixed operational pool $\mathcal{P}$ \newline
\textbf{Out:} Reconstructed global dense velocity field $\hat{\mathbf{u}}(\mathbf{x})$ &
$\hat{\mathbf{u}}(\mathbf{x}) = \mathcal{R}_{\psi}(\mathbf{U}_s, \mathbf{X}_s, g),\ \mathbf{X}_s \subset \mathcal{P}$ \newline $\forall \mathbf{x} \in \Omega$ \\ \midrule

\textbf{3. Airworthiness \newline Wind-Compliance Risk} &
Score a candidate trajectory against the worst exceedance of three operational wind / gust thresholds drawn from UAM operations literature, the FAR §23.341 / §25.341 design-gust criterion, and publicly-disclosed eVTOL prototype envelopes. &
\textbf{In:} 4D trajectory $\mathbf{r}(t)$; Wind environment tensor $\mathbf{E}(\mathbf{x}) = (\mathbf{u}(\mathbf{x}), k(\mathbf{x}))$ \newline
\textbf{Out:} Continuous risk score $R \in [0,1]$ &
$R = h_{\phi}(\mathbf{r}(t), \mathbf{E}(\mathbf{r}(t)))$ \\ \midrule

\textbf{4. Site Wind- \newline Exposure Ranking} &
Rank candidate rooftops by the LBM-LES 95th-percentile horizontal wind speed inside the approach cylinder above each site. &
\textbf{In:} Statistical feature vector $\Phi(c_i)$ for candidate rooftop $c_i \in \mathcal{C}$ \newline
\textbf{Out:} Wind-exposure ranking $\pi: \mathcal{C} \rightarrow \{1, \dots, |\mathcal{C}|\}$ &
$s_{gt}(c_i) = -\,U_{95}^{\mathrm{cyl}}(c_i)$ \\ \midrule

\textbf{5. Noise \newline Propagation} &
Predict the building-shadowed, wind-modulated ground SPL footprint and community noise exposure. &
\textbf{In:} Source pos. $\mathbf{x}_s$; Sound power $L_W$; Wind velocity $\mathbf{u}(\mathbf{x})$; TKE $k(\mathbf{x})$ \newline
\textbf{Out:} A-weighted sound pressure level (SPL) map $\hat{L}_{p,A}(\mathbf{x}_r)$ &
$\hat{L}_{p,A}(\mathbf{x}_r) = \mathcal{A}_{\omega}(\mathbf{x}_s, L_W, \mathbf{u}, k, g)$ \\
\bottomrule
\end{tabular}
\end{table}

\subsection{Evaluation Protocols and Data Splits}
\label{subsec:protocols}
Shanghai is designated as the primary benchmark city due to its complex urban morphology, including dense high-rise clusters and prominent river corridors that induce strong direction-dependent wake patterns. To rigorously assess the out-of-distribution generalization of the models, we enforce a strict independent-direction (IE) split. Specifically, simulations for inflow azimuths of $90^{\circ}$ and $270^{\circ}$ are held out exclusively for testing, while the remaining 14 directions are utilized for training and validation. Each held-out direction is bracketed by trained neighbours $22.5^\circ$ away, so the IE split probes directional generalization between distinct inflow states. For the cross-city (CC) hold-out used in Section~\ref{subsec:ccgen} and \ref{app:task4}, all cases of the held-out city (Hong Kong) form the test set, and the Shanghai corpus is partitioned into training and validation at an 85/15 case ratio.

The performance for each task is quantified using the following mathematical metrics:

\begin{enumerate}
    \item \textbf{Volume-Integrated Relative $L_2$ Error ($\epsilon_{L_2}$):} Used in Tasks 1 and 2 to evaluate global field reconstruction accuracy:
    \begin{equation}
        \epsilon_{L_2} = \frac{\|\hat{\mathbf{u}} - \mathbf{u}\|_2}{\|\mathbf{u}\|_2} = \frac{\sqrt{\sum_{i \in \Omega_{fluid}} |\hat{\mathbf{u}}_i - \mathbf{u}_i|^2}}{\sqrt{\sum_{i \in \Omega_{fluid}} |\mathbf{u}_i|^2}}
    \end{equation}
    where $\hat{\mathbf{u}}$ and $\mathbf{u}$ denote the predicted and ground-truth velocity fields, respectively.

    \item \textbf{Spectral Error ($\epsilon_E$):} Used in Tasks 1 and 2 to assess whether the predicted fields preserve coherent turbulent structures across spatial scales:
    \begin{equation}
        \epsilon_E = \int | \log E_{\hat{u}}(\kappa) - \log E_{u}(\kappa) | d\kappa
    \end{equation}
    where $E(\kappa)$ represents the horizontal kinetic energy spectrum at wavenumber $\kappa$.

    \item \textbf{Regression error on the continuous risk score:} Used in Task 3 as the primary scoring metric. We report mean absolute and root-mean-square error of the predicted risk $\hat r \in [0, 1]$ against the ground-truth label, together with the Spearman rank correlation $\rho_s$ for ordering fidelity:
    \begin{equation}
        \text{MAE}(\hat r) = \tfrac{1}{n}\sum_{i=1}^{n} |\hat r_i - r_i|,\quad
        \text{RMSE}(\hat r) = \sqrt{\tfrac{1}{n}\sum_{i=1}^{n} (\hat r_i - r_i)^2}.
    \end{equation}

    \item \textbf{Threshold-derived classification metrics:} Used in Task 3 alongside operator-defined go/no-go thresholds. The AUC-ROC, Brier score, and expected cost are inline-derived by thresholding the ground-truth label at $0.5$:
    \begin{equation}
        \text{AUC-ROC} = \int_{0}^{1} TPR(FPR^{-1}(t))\, dt,
    \end{equation}
    where $TPR$ and $FPR$ are the True Positive Rate and False Positive Rate of the thresholded binary problem. These metrics are reported with an ``\,@$0.5$'' suffix in Table~\ref{tab:task3} and are not the optimisation target of the task.

    \item \textbf{Normalized Discounted Cumulative Gain (NDCG@10):} Used in Task 4 to evaluate the accuracy of the top-10 ranked vertiport candidates:
    \begin{equation}
        \text{NDCG@k} = \frac{DCG_k}{IDCG_k}, \quad DCG_k = \sum_{i=1}^{k} \frac{2^{rel_i} - 1}{\log_2(i+1)}
    \end{equation}
    where $rel_i$ is the relevance score of the candidate at rank $i$, and $IDCG$ is the ideal $DCG$.
\end{enumerate}

\section{Experimental Evaluation\label{sec:4}}

While the \textsc{U3DWind} dataset encompasses five distinct cities, the benchmark experiments presented in this study primarily utilize Shanghai as the reference domain. This selection ensures a reproducible and rigorous evaluation against a highly complex urban morphology. Specifically, the Shanghai domain features dense high-rise clusters, the Huangpu River corridor, and pronounced direction dependent wake patterns. These characteristics are highly representative of the low-altitude operational environments for UAM in major metropolitan areas.

\subsection{Task 1: Surrogate Modeling}

The quantitative results for Task 1, as summarized in Table~\ref{tab:task1}, reveal distinct performance tiers across different modeling paradigms. Throughout, SSIM is computed on the de-normalized velocity magnitude (m/s), per horizontal slice with an $11 \times 11$ Gaussian window ($\sigma = 1.5$) restricted to fluid cells, using the per-case dynamic range $L$ of the reference field ($C_1 = (0.01L)^2$, $C_2 = (0.03L)^2$), and averaged over slices with fluid-empty slices excluded. The learned rows train high-capacity models on at most 125 Shanghai cases (train plus validation); this small-sample regime is mitigated by the 40~m downsampled inputs, the multi-seed protocol, and the directional hold-out.
\begin{table}[t]
\centering
\caption{Task 1 baseline matrix on the Shanghai \textsc{U3DWind} release. Evaluation is on the IE split, which holds out two inflow directions ($90^\circ$ and $270^\circ$). Inputs are four-times-downsampled fields at 40\,m effective resolution with a $384 \times 384 \times 32$ central crop. Rows with $\pm$ report the mean and standard deviation over three training seeds; the remaining rows are single-seed runs. The $\sigma_{\mathrm{UQ}}$ column reports the wrapper-specific uncertainty estimate: the calibrated predictive standard deviation for the MC-dropout variant and the split-conformal radius $q_{\alpha=0.1}$ (in m/s) for the conformal Swin-3D row; the member standard deviation for the two deep-ensemble rows. Best value per column is bold; the non-converged denoising-diffusion row is excluded from the bolding. $^{a}$Trainable parameters, counted from the released code as instantiated by the runners; $^{b}$under neuraloperator~2.0; $^{c}$XGBoost capacity is 800 trees of depth $\le 6$ per run, so no dense parameter count applies; $^{d}$trainable gate only, on top of frozen UNet and FNO backbones (6.66\,M frozen).}
\label{tab:task1}
\footnotesize
\setlength{\tabcolsep}{4pt}
\begin{tabular}{lcccccc}
\toprule
Baseline & Params$^{a}$ & $\varepsilon_{L_2}\downarrow$ & SSIM $\uparrow$ & slab MAE $\downarrow$ & $\varepsilon_E\downarrow$ & $\sigma_{\mathrm{UQ}}$ \\
\midrule
Log-law projection       & --- & $1.4960$                 & $0.7306$                 & $0.9760$                 & $1.4130$                 & --- \\
3D kriging               & --- & $0.6423$                 & $0.6735$                 & $0.2939$                 & $\mathbf{1.1587}$        & --- \\
XGBoost per voxel        & ---$^{c}$ & $0.6620$                 & $0.7817$                 & $0.3378$                 & $1.4223$                 & --- \\
3D-CNN (no skip)         & 0.53\,M & $0.6078$                 & $0.6998$                 & $0.3489$                 & $1.6277$                 & --- \\
3D UNet                  & 1.40\,M & $0.5909\pm0.0662$        & $0.7503\pm0.0168$        & $0.3062\pm0.0235$        & $1.4449\pm0.0959$        & --- \\
3D ResUNet               & 2.14\,M & $0.6013\pm0.0332$        & $0.7554\pm0.0094$        & $0.3294\pm0.0100$        & $1.4453\pm0.0561$        & --- \\
Attention UNet           & 1.42\,M & $0.5980\pm0.0390$        & $0.7566\pm0.0254$        & $0.3282\pm0.0284$        & $1.3812\pm0.1141$        & --- \\
FNO-3D                   & 5.26\,M$^{b}$ & $0.6877\pm0.0196$        & $0.7776\pm0.0015$        & $0.2470\pm0.0040$        & $1.5615\pm0.0074$        & --- \\
DeepONet                 & 0.18\,M & $0.5237\pm0.0367$        & $\mathbf{0.7819\pm0.0028}$ & $0.2433\pm0.0113$      & $1.5555\pm0.0434$        & --- \\
Swin-3D                  & 0.30\,M & $\mathbf{0.5042\pm0.0040}$ & $0.7296\pm0.0072$      & $0.2609\pm0.0101$        & $1.5883\pm0.0554$        & --- \\
PINN-UNet                & 1.40\,M & $0.5508\pm0.0158$        & $0.7648\pm0.0064$        & $0.2963\pm0.0280$        & $1.5405\pm0.0603$        & --- \\
Multi-fidelity ensemble  & 0.013\,M$^{d}$ & $0.6887$                 & $0.7780$                 & $0.2492$                 & $1.5481$                 & --- \\
Denoising diffusion (3D)  & 1.41\,M & $9.1502$                 & $0.1103$                 & $6.0009$                 & $1.1355$                 & --- \\
\midrule
GINO                     & 5.26\,M$^{b}$ & $0.6730$                 & $0.7808$                 & $0.2435$                 & $1.5493$                 & --- \\
\midrule
UNet-3D deep ensemble ($M=5$)      & 7.02\,M & $0.5164$       & $0.7711$                 & $0.2593$                 & $1.4986$ & $0.1651$ \\
UNet-3D MC-dropout ($K=20, p=0.1$) & 1.40\,M & $0.5925$       & $0.7453$                 & $0.3316$                 & $1.4320$ & $0.3769$ \\
FNO-3D deep ensemble ($M=3$)       & 15.77\,M$^{b}$ & $0.6799$       & $0.7802$                 & $\mathbf{0.2409}$        & $1.5815$ & $0.0843$ \\
Swin-3D conformal ($\alpha=0.1$)   & 0.30\,M & $0.5068$       & $0.7301$                 & $0.2566$                 & $1.6008$ & $3.2833$ \\
\bottomrule
\end{tabular}%
\end{table}
The evaluation on the Shanghai IE split underscores the inherent complexity of urban micro-climatology and the varying capacities of neural architectures to resolve building-induced flow features.

The analytical Log-law projection yields a prohibitive relative error ($\varepsilon_{L_2} = 1.4960$), the largest in the matrix, together with the highest slab MAE ($0.9760$). Although its Structural Similarity Index (SSIM) of $0.7306$ appears moderate, this score stems from the smooth vertical stratification shared with the reference: conventional logarithmic wind profiles remain fundamentally incapable of capturing blockage effects, downdrafts, and canyon-jet accelerations induced by heterogeneous urban morphologies. \textit{3D Kriging} reaches $\varepsilon_{L_2} = 0.6423$ and, notably, the lowest spectral error of all baselines ($\varepsilon_E = 1.1587$), yet its SSIM of $0.6735$ is the lowest in the matrix, indicating that linear statistical interpolation over-smooths the sharp gradients at fluid-solid interfaces even while preserving the bulk energy distribution.

Learned models demonstrate a significant performance improvement over the analytical baseline. \textit{Swin-3D} achieves the best global accuracy with $\varepsilon_{L_2} = 0.5042 \pm 0.0040$, and its small seed spread makes it the most stable learned baseline on this split. \textit{DeepONet} \citep{Lu2021DeepONet} follows at $\varepsilon_{L_2} = 0.5237 \pm 0.0367$ while attaining the highest structural fidelity of the matrix ($\text{SSIM} = 0.7819 \pm 0.0028$) and the lowest Mean Absolute Error (MAE) among the deterministic baselines ($0.2433 \pm 0.0113$) within the UAM-critical altitude slab ($55.3 \le z \le 295.9$~m), suggesting that the branch-trunk architecture effectively decouples global inflow conditions from local geometric features. \textit{FNO-3D} \citep{Li2021FNO,Wen2022UFNO}, by contrast, records $\varepsilon_{L_2} = 0.6877 \pm 0.0196$, near the bottom of the learned family, although its slab MAE of $0.2470 \pm 0.0040$ remains competitive; the spectral parameterization evidently preserves smooth large-scale transport patterns \citep{Pathak2022FourCastNet} without resolving the held-out inflow directions probed by the IE split.

A critical observation is the decoupling between integrated spatial errors and spectral fidelity. The lowest spectral error among converged baselines is attained by \textit{3D Kriging} ($\varepsilon_E = 1.1587$), a smooth estimator with modest point accuracy, which shows that $\varepsilon_E$ rewards preservation of the bulk energy distribution independently of structural correctness. Within the learned family, the skip-connected UNet family retains lower spectral errors (\textit{Attention UNet} at $1.3812 \pm 0.1141$ and \textit{3D UNet} at $1.4449 \pm 0.0959$) than the neural operators (\textit{DeepONet} at $1.5555 \pm 0.0434$, \textit{FNO-3D} at $1.5615 \pm 0.0074$, and \textit{Swin-3D} at $1.5883 \pm 0.0554$), indicating that the operators' accuracy gains come with a measurable spectral bias \citep{Raissi2020Hidden}. Notably, \textit{PINN-UNet} \citep{Raissi2019PINN,Cai2021PINNFluid} shows no spectral advantage: its $\varepsilon_E = 1.5405 \pm 0.0603$ exceeds that of the purely data-driven \textit{3D UNet}, so under the current training budget the continuity-residual regularizer improves the integrated error ($\varepsilon_{L_2} = 0.5508 \pm 0.0158$, third best among deterministic baselines), leaving the spectral error at the level of the neural operators.

Despite the advancements in magnitude prediction, all models plateau at an SSIM below $0.79$, with the learned baselines spanning $0.6998$ to $0.7819$. The remaining headroom concentrates in high-gradient flow features, such as separation bubbles and recirculation zones in the immediate lee of high-rise structures. At the effective resolution of 40~m, baseline models tend to yield over-smoothed predictions in regions dominated by complex wake interactions, suggesting that further research into multi-scale loss functions or generative refinements is required.

The inclusion of uncertainty-aware variants provides further insights into model robustness. The \textit{UNet-3D deep ensemble} ($M = 5$) lowers the integrated error of its deterministic base from $0.5909$ to $0.5164$, the strongest accuracy gain in the wrapper family, although the ensemble rows report the epistemic spread of their members ($0.1651$ for the UNet-3D ensemble, $0.0843$ for the FNO-3D ensemble). The \textit{FNO-3D deep ensemble} ($M = 3$) behaves similarly, moving the FNO-3D error from $0.6877$ to $0.6799$ while attaining the lowest slab MAE of the entire matrix ($0.2409$). The \textit{UNet-3D MC-dropout} variant (with $K = 20$ stochastic forward passes and $p = 0.1$) reports a mean prediction at $\varepsilon_{L_2} = 0.5925$, essentially matching the deterministic UNet, together with a calibrated predictive standard deviation of $\sigma_{\mathrm{UQ}} = 0.3769$ that the deterministic model does not provide. The \textit{Swin-3D conformal} wrapper ($\alpha = 0.1$) emits a split-conformal radius of $q_\alpha = 3.2833$\,m/s on top of a point prediction that closely tracks the base Swin-3D ($\varepsilon_{L_2} = 0.5068$ versus $0.5042$). With six calibration cases the attainable miscoverage level is $\alpha_{\mathrm{eff}} = 0.1429$, and the measured case-level coverage on the held-out directions is $1.0$; because the calibration cases carry trained inflow directions while the test cases are the held-out $90^\circ$ and $270^\circ$ directions, this coverage is an empirically measured rate. Together, these variants give downstream UAM risk assessment a reliable aerodynamic input with post-hoc uncertainty estimates: a calibrated $\sigma_{\mathrm{UQ}}$ from MC-dropout and a conformal coverage radius for the Swin-3D wrapper. A denoising-diffusion surrogate trained under the same shared budget failed to converge ($\varepsilon_{L_2} = 9.15$, SSIM $0.11$) and is included in Table~\ref{tab:task1}; generative surrogates require training budgets or conditioning beyond the shared protocol used here.

\begin{figure}
    \centering
    \includegraphics[width=\textwidth]{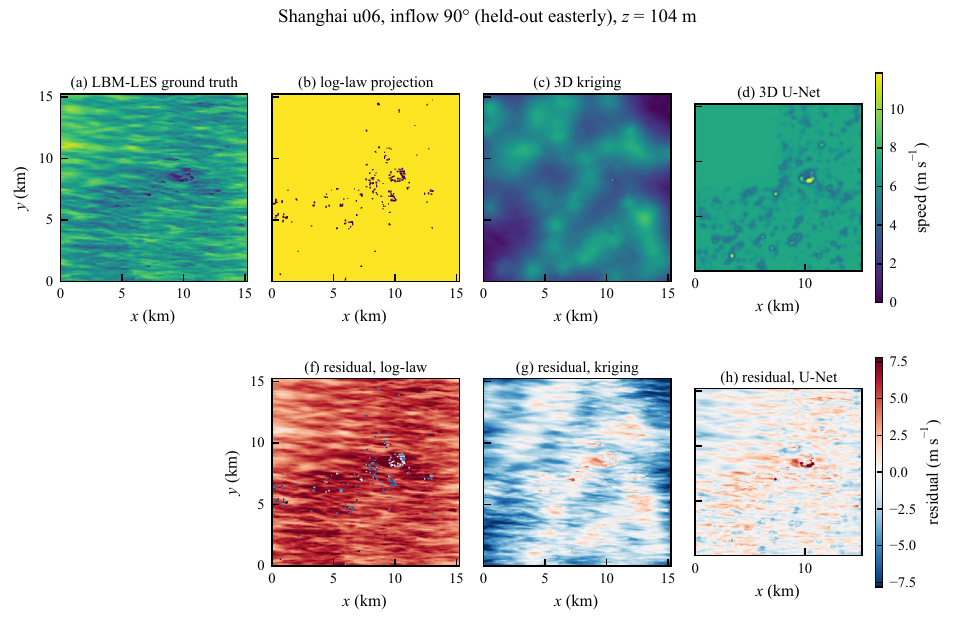}
    \caption{Task 1 qualitative prediction at $z=104$\,m on the held-out easterly-inflow case (u06, inflow $90^\circ$) over the centered $384 \times 384$-cell crop. Panel a is the LBM-LES ground truth; panels b, c, d show representative members of the three baseline families: the analytical log-law projection is horizontally uniform by construction and over-predicts the canopy-decelerated flow at this height, kriging produces a blurred field that captures broad corridors but misses wake detail, and the three-dimensional U-Net stays close to a smooth conditional mean in sparsely built sectors while resolving local wakes around the resolved buildings. Panels f, g, h give the signed residuals on a shared scale: the log-law projection over-shoots by up to $8$\,m/s, kriging under-predicts most of the slice, and the U-Net residuals are the smallest overall but concentrate in the high-rise cluster wake.}
    \label{fig:task1_slice}
\end{figure}

The visual comparison of the horizontal speed field at $z=104$\,m (Fig.~\ref{fig:task1_slice}) makes the capacities of the three baseline families concrete. The \textit{LBM-LES ground truth} (Fig.~\ref{fig:task1_slice}a) is a heterogeneous field of building-induced wakes and localized channelization. The analytical \textit{Log-law projection} (Fig.~\ref{fig:task1_slice}b) is horizontally uniform by construction, over-predicts the canopy-decelerated flow at this height, and over-shoots by up to $8$\,m\,s$^{-1}$ in the wake regions (Fig.~\ref{fig:task1_slice}f). The \textit{3D Kriging} baseline (Fig.~\ref{fig:task1_slice}c) recovers broad corridors but under-predicts most of the slice and smears the high-gradient wake boundaries of individual high-rise clusters (Fig.~\ref{fig:task1_slice}g). The \textit{3D U-Net} (Fig.~\ref{fig:task1_slice}d) resolves local wakes around the resolved buildings while staying close to a smooth conditional mean in sparsely built sectors; its residuals (Fig.~\ref{fig:task1_slice}h) are the smallest overall but concentrate in the high-rise cluster wake, the flow structure hardest to internalize from the training directions.

\subsection{Task 2: Sparse Reconstruction}

The experimental results for Task 2, as detailed in Tables~\ref{tab:task2_recon} and~\ref{tab:task2_sensor_sweep}, evaluate the capacity of various baseline paradigms to recover high-resolution volumetric wind fields from a limited set of rooftop and corridor observations \citep{Fukami2021Voronoi}. The results highlight the critical role of physically-structured priors in addressing the ill-posed nature of urban wind field reconstruction \citep{Zhang2022WindPINN}.

The results show that reduced-order models leveraging a dataset-derived prior remain the most accurate option at $N_s = 64$, although their margin over the strongest sensor-conditioned neural baseline has narrowed considerably. \textit{Kalman assimilation} (\textsc{U3DWind} prior) \citep{Anderson2001EnKF} attains the lowest errors, with a relative $L_2$ error of $0.142$ and a slab MAE of $0.674$, while \textit{Gappy POD} \citep{Willcox2006Gappy} is nearly tied with it at $\varepsilon_{L_2} = 0.143$ and attains the highest spectral fidelity ($\varepsilon_E = 0.116$), successfully recovering the dominant modal structures of the urban wake with 64 sensors. This suggests that the latent space of the \textsc{U3DWind} dataset effectively captures the coherent flow features of the urban boundary layer, allowing assimilation methods to ``fill in'' the unobserved street canyons by projecting sparse measurements onto physically consistent modes.

Among the neural-based baselines, the \textit{Sensor Transformer} (using query-grid attention) is the most competitive architecture, achieving $\varepsilon_{L_2} = 0.146$ and trailing the assimilation pair by less than $0.005$. A second tier formed by the \textit{Physics-informed reconstruction} ($0.186$), the \textit{Graph-attention reconstruction} ($0.188$), and the \textit{Masked 3D-UNet} ($0.198$) also recovers most of the volumetric field. The Sensor Transformer and the graph-attention model encode the sensor cloud directly as positions and values without scattering it onto a near-empty 3D grid, which gives them a non-trivial inductive bias when only about 250 IE training cases are available. A score-based conditional diffusion baseline run under the same budget did not converge ($\varepsilon_{L_2} = 1.68$) and is reported in Table~\ref{tab:task2_recon}. The remaining deep baselines degrade more visibly, with the \textit{Normalizing-flow posterior} at $0.310$, \textit{FNO-3D with sensor tokens} at $0.447$, and the \textit{Voronoi CNN} at $0.586$, all of which nevertheless remain well below the $\varepsilon_{L_2} \approx 1.0$ error of regressing the dataset-mean field. The IE split nonetheless remains a data-scarce regime: the fluid-cell coverage of the \textsc{U3DWind} LBM-LES output is large (around $31 \times 623 \times 748$ voxels on the merged two-city grid) and the IE split supplies only a few hundred training cases, so deep parametric models without strong sensor-aware structural priors still trail the reduced-order methods in spectral fidelity and sensor scaling.

The sensor-efficiency sweep in Table~\ref{tab:task2_sensor_sweep} reveals that the two families scale in opposite ways. The prior-based methods convert additional sensors into accuracy: \textit{Kalman assimilation} improves from $\varepsilon_{L_2} = 0.189$ at $N_s=16$ to $0.128$ at $N_s=256$, and \textit{Gappy POD} from $0.221$ to $0.128$. The neural baselines are instead nearly flat in sensor count: the \textit{Sensor Transformer} moves only from $0.151$ to $0.144$, and the \textit{Graph-attention reconstruction} degrades from $0.178$ to $0.204$. As a consequence, the Sensor Transformer is the single best method in the sparsest regime ($0.151$ versus $0.189$ for Kalman assimilation at $N_s=16$), while the assimilation pair overtakes it from $N_s=64$ onwards. Deep models are sample-efficient at low sensor counts but fail to exploit denser instrumentation, whereas methods that explicitly incorporate the dataset's low-dimensional physical structure continue to improve as sensors are added.

The spectral error ($\varepsilon_E$) serves as a discriminator for the operational utility of the reconstruction and separates methods that the $L_2$ metric ranks closely. The assimilation-based methods preserve the energy content best, with $\varepsilon_E = 0.116$ for \textit{Gappy POD} and $0.120$ for \textit{Kalman assimilation}, followed at a distance by classical interpolants such as \textit{3D kriging} ($0.269$) and \textit{Voronoi nearest-neighbor} ($0.273$). The \textit{Sensor Transformer}, in contrast, combines competitive $L_2$ accuracy with one of the largest spectral errors ($0.454$), indicating that its reconstructions are spectrally distorted even where they are pointwise accurate. Since preserving the spatial spectral content and spatial heterogeneity of the urban wind field is essential for safety-critical UAM operations such as localized gust estimation and vertiport approach monitoring, the dataset-derived reduced-order prior remains clearly preferable on this axis.

\begin{table}[htbp]
\centering
\caption{Task 2 sparse-sensor reconstruction at $N_s = 64$ sensors on the merged Shanghai and Hong Kong \textsc{U3DWind} corpus (resampled to a common grid; 36 IE test cases), evaluated on the IE split (held-out inflow directions of $90^\circ$ and $270^\circ$). Entries are means over three random seeds (two for kriging; single seed for the conditional diffusion row, which is excluded from the per-column bolding). The sensor placement is drawn from the deterministic operational sensor pool $\mathcal{P}$ spanning rooftop, corridor-anchor, and mixed-tier candidates. Best value per column is bold.}
\label{tab:task2_recon}
\footnotesize
\begin{tabular}{lccc}
\toprule
Baseline & $\varepsilon_{L_2}\downarrow$ & slab MAE $\downarrow$ & $\varepsilon_E\downarrow$ \\
\midrule
Inverse-distance weighting                 & $0.672$          & $1.421$          & $0.362$          \\
Voronoi nearest-neighbor                  & $0.540$          & $1.150$          & $0.273$          \\
Thin-plate-spline RBF                      & $1.127$          & $1.284$          & $0.337$          \\
3D kriging (Mat\'ern-5/2)                  & $0.552$          & $1.437$          & $0.269$          \\
Gappy POD                                  & $0.143$          & $0.677$          & $\mathbf{0.116}$ \\
Kalman assimilation (\textsc{U3DWind} prior)        & $\mathbf{0.142}$ & $\mathbf{0.674}$ & $0.120$          \\
Voronoi CNN                                & $0.586$          & $1.735$          & $0.468$          \\
Masked 3D-UNet                             & $0.198$          & $0.962$          & $0.350$          \\
FNO-3D with sensor tokens                  & $0.447$          & $1.798$          & $0.371$          \\
Sensor Transformer (query-grid attention)  & $0.146$          & $0.724$          & $0.454$          \\
Graph-attention reconstruction             & $0.188$          & $0.824$          & $0.322$          \\
Physics-informed reconstruction            & $0.186$          & $0.909$          & $0.360$          \\
Normalizing-flow posterior                 & $0.310$          & $1.230$          & $0.368$          \\
Conditional diffusion (single seed)        & $1.676$          & $5.203$          & $0.460$          \\
\bottomrule
\end{tabular}
\end{table}

\begin{table}[htbp]
\centering
\caption{Task 2 sensor-efficiency sweep on the merged Shanghai and Hong Kong corpus. Relative $L_2$ error $\varepsilon_{L_2}$ at increasing sensor counts $N_s \in \{16, 64, 256\}$ on the IE split; smaller is better. Entries are means over three random seeds (two for kriging). Best value per column is bold; at $N_s = 256$ Kalman assimilation and Gappy POD are tied at the reported precision.}
\label{tab:task2_sensor_sweep}
\footnotesize
\begin{tabular}{lccc}
\toprule
Baseline & $N_s=16$ & $N_s=64$ & $N_s=256$ \\
\midrule
Inverse-distance weighting           & $0.872$          & $0.672$          & $0.531$ \\
Voronoi nearest-neighbor            & $0.562$          & $0.540$          & $0.526$ \\
Thin-plate-spline RBF                & $1.336$          & $1.127$          & $1.200$ \\
3D kriging                           & $0.626$          & $0.552$          & $0.538$ \\
Gappy POD                            & $0.221$          & $0.143$          & $\mathbf{0.128}$ \\
Kalman assimilation                  & $0.189$          & $\mathbf{0.142}$ & $\mathbf{0.128}$ \\
Voronoi CNN                          & $0.598$          & $0.586$          & $0.593$ \\
Masked 3D-UNet                       & $0.204$          & $0.198$          & $0.201$ \\
FNO-3D with sensor tokens            & $0.459$          & $0.447$          & $0.577$ \\
Sensor Transformer                   & $\mathbf{0.151}$ & $0.146$          & $0.144$ \\
Graph-attention reconstruction       & $0.178$          & $0.188$          & $0.204$ \\
\bottomrule
\end{tabular}
\end{table}

\begin{figure}
    \centering
    \includegraphics[width=\linewidth]{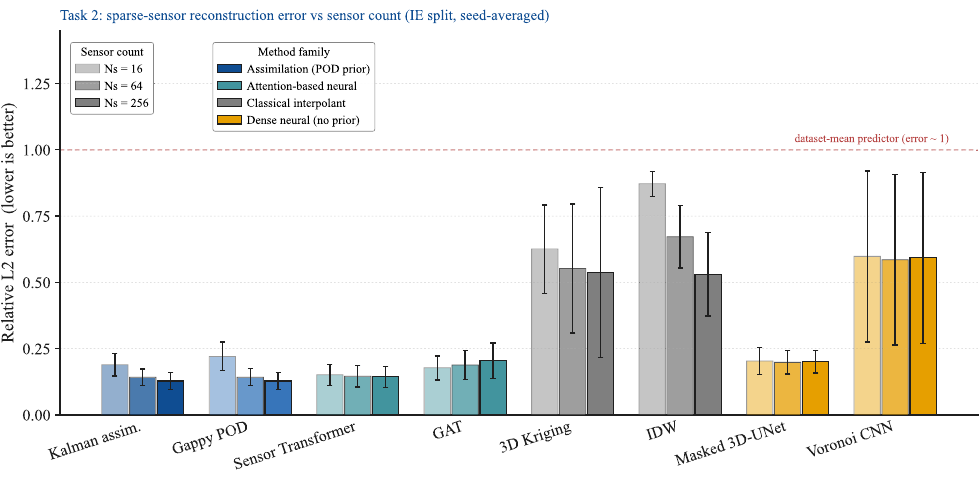}
    \caption{Task 2 sensor-count scaling. Relative $L_2$ reconstruction error on the held-out inflow directions as a function of the number of rooftop sensors $N_s$. Kalman assimilation and Gappy POD improve steadily with sensor count and lead for $N_s \geq 64$; the sensor Transformer is the most accurate method at $N_s = 16$ but stays nearly flat as sensors are added, and the graph-attention baseline degrades; classical spatial interpolants (inverse-distance weighting, kriging) never close the gap.}
    \label{fig:task2_scaling}
\end{figure}

The scaling behavior illustrated in Fig.~\ref{fig:task2_scaling} highlights the distinct convergence characters of physics-prior-driven assimilation and purely data-driven neural architectures. Methods leveraging a dataset-derived reduced-order prior, specifically \textit{Kalman assimilation} and \textit{Gappy POD}, improve monotonically with sensor count, realizing most of their gain between $N_s=16$ and $N_s=64$ and a further modest reduction to $\varepsilon_{L_2} = 0.128$ at $N_s=256$. The attention-based neural baselines behave differently: the \textit{Sensor Transformer} is already the most accurate method at $N_s=16$ but remains essentially flat thereafter, and the \textit{Graph-attention (GAT)} model worsens as sensors are added, so both are overtaken by the assimilation family at $N_s \geq 64$. Convolutional and spectral baselines such as \textit{Voronoi CNN}, \textit{Masked UNet}, and \textit{FNO-3D} likewise show little benefit from denser instrumentation, underscoring the difficulty of processing highly sparse, unstructured meteorological observations with standard grid-based operators. These results indicate that for operational UAM monitoring, physically-informed priors are what allow additional sensing infrastructure to be translated into higher-fidelity situational awareness.

\begin{figure}[t]
    \centering
    \includegraphics[width=\linewidth]{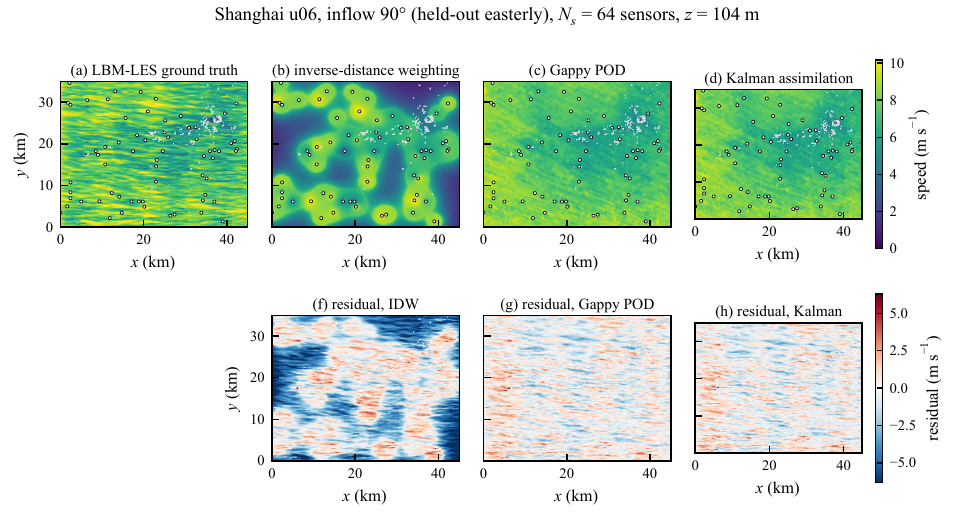}
    \caption{Task 2 reconstruction at $z=104$\,m on the held-out easterly-inflow case (Shanghai, u06) with $N_s = 64$ sensors drawn from the mixed rooftop-and-corridor placement pool (white markers, projected onto the plotted plane). Panel a is the LBM-LES ground truth; panels b, c, d are three reconstructions spanning the methodological axis: inverse-distance weighting is a geometry-only baseline that produces blob-like interpolants unable to capture building wakes, while Gappy POD and Kalman assimilation, both operating in the same $K=24$-mode POD subspace, recover the dominant urban channels and are nearly indistinguishable at this sensor density (volume relative $L_2$ of $0.10$ for both on this case). Panels f, g, h show the signed reconstruction errors on a shared scale: the projection-based methods leave streak-scale residuals of about $\pm 3$\,m/s that 64 sensors cannot pin down, structure an order of magnitude finer than the IDW error field.}
    \label{fig:task2_recon}
\end{figure}

The qualitative fidelity of wind field reconstruction under sparse sensing ($N_s = 64$) is illustrated in Fig.~\ref{fig:task2_recon}. The comparison across different reconstruction paradigms highlights the necessity of incorporating physical priors to resolve complex urban flow structures:

\begin{itemize}
    \item \textbf{Heuristic Interpolation:} The \textit{Inverse-distance weighting} (IDW) baseline (Fig.~\ref{fig:task2_recon}b) serves as a geometry blind reference. It produces a highly localized and fragmented field where ``blobs'' of velocity are centered strictly around sensor locations. The corresponding error map (Fig.~\ref{fig:task2_recon}f) reveals extensive residuals across the domain, as IDW is fundamentally incapable of inferring building-induced wakes or channeled flows in unsensed regions.

    \item \textbf{Reduced-Order Representation:} \textit{Gappy POD} with $K=24$ modes (Fig.~\ref{fig:task2_recon}c) demonstrates a significant improvement by projecting sparse observations onto a physically-consistent latent space. It recovers the primary urban channels and macroscopic wake orientations that are absent in the IDW baseline, while the projection leaves streak-scale residuals of about $\pm 3$\,m\,s$^{-1}$ in the error map (Fig.~\ref{fig:task2_recon}g), structure that 64 sensors cannot pin down.

    \item \textbf{Optimal Data Assimilation:} \textit{Kalman assimilation}, operating in the same $K=24$-mode POD subspace (Fig.~\ref{fig:task2_recon}d), combines the reduced-order prior with the sensor observations through an optimal gain and is nearly indistinguishable from Gappy POD at this sensor density (volume relative $L_2$ of $0.10$ for both on this case). Its residual map (Fig.~\ref{fig:task2_recon}h) exhibits the same streak-scale structure, with remaining discrepancies localized around high-turbulence building corners where sub-grid effects are most prominent.
\end{itemize}

This qualitative hierarchy confirms that for UAM situational awareness, data assimilation methods leveraging the \textsc{U3DWind} dataset's physical manifold provide a robust mechanism for inferring the complete volumetric state from limited rooftop instrumentation.

\subsection{Cross-City Generalization}
\label{subsec:ccgen}

The four representative Task 1
architectures are additionally evaluated under a cross-city hold-out: the
models are trained on the Shanghai corpus and tested on all 144 Hong Kong
cases, with three seeds per architecture and the same metric definitions as
the IE tables. Table~\ref{tab:task1_cc} pairs the in-city and cross-city
scores. Transfer costs a consistent $0.19$ to $0.22$ increase in relative
$L_2$ (31 to 42\,\%) and an SSIM drop of $0.07$ to $0.11$, but the $\varepsilon_{L_2}$ architecture ordering is preserved exactly across the two settings
(DeepONet $<$ PINN-UNet $<$ UNet-3D $<$ FNO-3D on both splits), so model selection performed in-city carries over to the unseen city.

\begin{table}[htbp]
\centering
\caption{Task 1 cross-city generalization: trained on Shanghai, tested on all
144 Hong Kong cases (mean $\pm$ sample std over three seeds; same metric
version as the IE tables). The large FNO-3D spread comes from one seed that
deterministically lands in a poor basin and fails to beat the zero-predictor
baseline; it is reported as is.}
\label{tab:task1_cc}
\footnotesize
\begin{tabular}{lcccc}
\toprule
Model & IE $\varepsilon_{L_2}$ & CC $\varepsilon_{L_2}$ & CC SSIM & CC slab MAE \\
\midrule
DeepONet  & $0.524 \pm 0.037$ & $\mathbf{0.743 \pm 0.003}$ & $\mathbf{0.707}$ & $\mathbf{0.444}$ \\
PINN-UNet & $0.551 \pm 0.016$ & $0.765 \pm 0.034$ & $0.685$ & $0.477$ \\
UNet-3D   & $0.591 \pm 0.066$ & $0.776 \pm 0.013$ & $0.657$ & $0.474$ \\
FNO-3D    & $0.688 \pm 0.020$ & $0.912 \pm 0.115$ & $0.671$ & $0.505$ \\
\bottomrule
\end{tabular}
\end{table}

\subsection{Physical Analysis}
\label{subsec:physics}

This subsection examines the physical content of the released fields: where urban wind hazards arise, how they are distributed over the UAM operating band, and which flow regimes defeat which model families.

\begin{figure}[htbp]
\centering
\includegraphics[width=\linewidth]{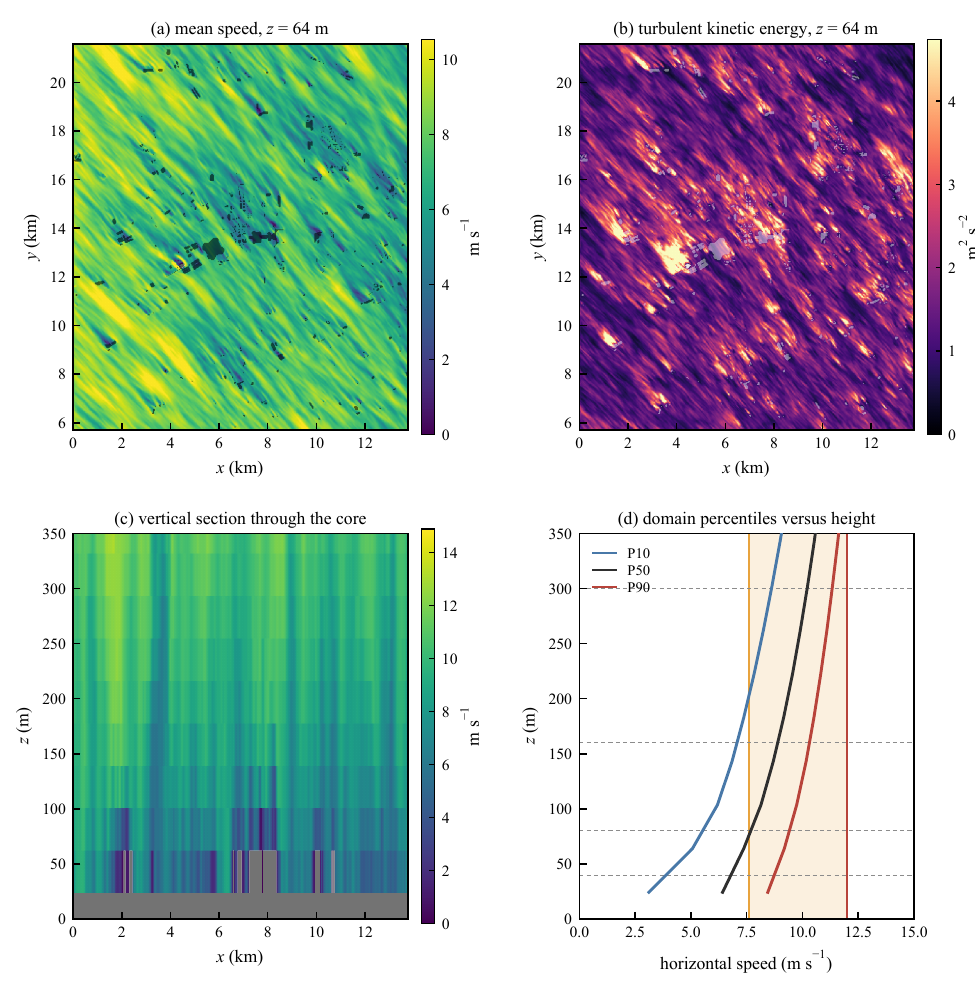}
\caption{Flow anatomy of the Shanghai urban core (annual u06 tier,
$135^\circ$ inflow, 40\,m archival fields). (a) Mean horizontal speed and
(b) TKE at $z=64$\,m with building footprints overlaid; (c) vertical section
through the core row; (d) domain percentiles of horizontal speed versus
height with the 7.6 and 12\,m/s thresholds and the UAM altitude bands marked.}
\label{fig:phys_atlas}
\end{figure}

Figure~\ref{fig:phys_atlas} maps the flow anatomy of the Shanghai urban core
(annual u06 tier, $135^\circ$ inflow, 40\,m archival fields). The plan views at
$z = 64$\,m show the three hazard mechanisms that recur across the corpus:
velocity deficits extending kilometers downwind of the supertall cluster,
turbulence hot spots hugging the flanks and rooftops of the tallest buildings,
and channeled acceleration along street and river alignments oriented with the
inflow. The vertical section through the core row shows the rooftop shear layer
that feeds these hot spots and the slow recovery of the wake with height. The
percentile profiles in panel (d) place this variability in operational terms:
across the UAM altitude bands the interdecile spread of horizontal speed reaches from well below the 7.6\,m/s sustained-wind threshold to above $11$\,m/s near the top of the operating band, so a go/no-go decision depends on where in the city a vehicle
flies, and a single city-average inflow statistic cannot substitute for the
resolved field.

\begin{figure}[htbp]
\centering
\includegraphics[width=0.54\linewidth]{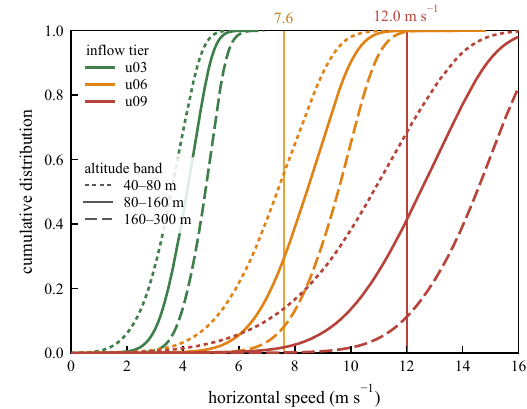}\hfill
\includegraphics[width=0.44\linewidth]{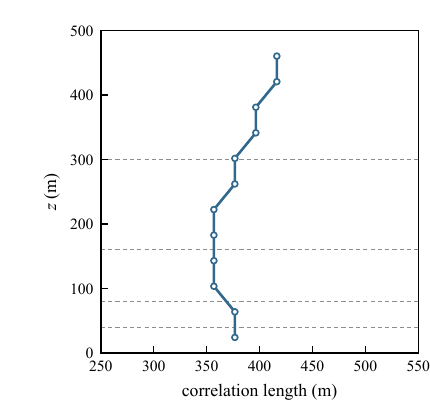}
\caption{Corpus statistics of the Shanghai release. Left:
altitude-band speed CDFs for the three annual inflow tiers ($135^\circ$
inflow) against the 7.6 and 12\,m/s thresholds. Right: two-point correlation
length of horizontal speed in the urban core versus height.}
\label{fig:phys_stats}
\end{figure}

Figure~\ref{fig:phys_stats} summarizes the same corpus statistically.
The altitude-band speed CDFs shift monotonically with inflow tier and with
altitude band, and the fraction of airspace above the sustained-wind threshold
grows steeply from the 40--80\,m band to the 160--300\,m band, which quantifies
the classical UAM trade of flying higher into stronger but smoother wind.
The two-point correlation length of horizontal speed in the urban core is
355--375\,m through the operating band. This scale carries two design
consequences: kilometer-scale mesoscale products cannot represent the hazard
structure, which motivates the decameter resolution of the released fields,
and a sensor network intended to constrain the core flow (Task~2) needs
spacing comparable to this correlation length to be informative.

\begin{figure}[htbp]
\centering
\includegraphics[width=\linewidth]{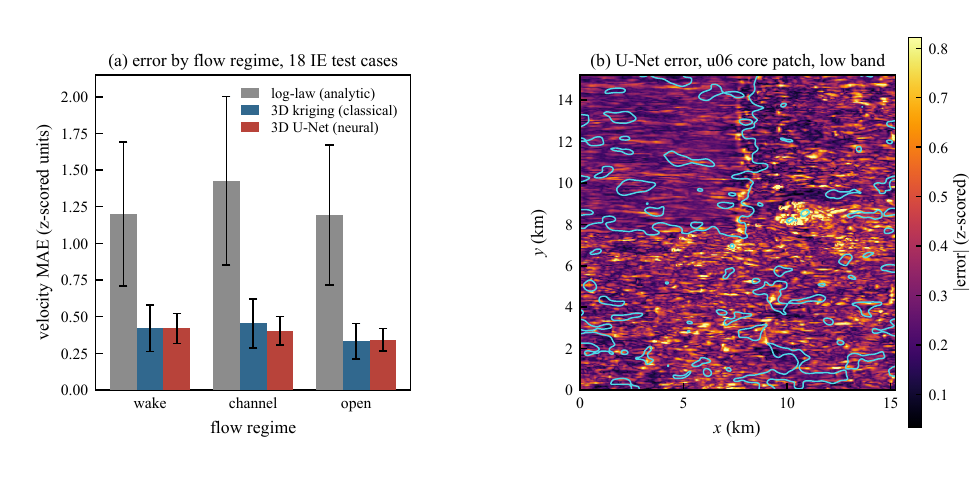}
\caption{Error attribution by flow regime over the 18 Shanghai IE
test cases. (a) Velocity MAE (z-scored) of an analytic, a classical, and a
neural baseline in the wake (TKE $>$ Q75), channel ($U >$ Q75, TKE $\le$ Q50),
and open regimes; regimes are computed per case from the reference fields.
(b) UNet error map over the urban core in the lowest altitude band with the
smoothed wake boundary overlaid.}
\label{fig:phys_regime}
\end{figure}

Figure~\ref{fig:phys_regime} attributes benchmark error to flow regimes.
For each of the 18 Shanghai IE test cases we partition the fluid volume,
using the reference fields only, into a wake regime (TKE above its 75th
percentile), a channel regime (speed above its 75th percentile with TKE below
the median), and open flow (the remainder), and evaluate one representative of
each model family. The analytic log-law profile fails in all regimes (z-scored
velocity MAE 1.20 wake / 1.43 channel / 1.19 open): it misses horizontal
structure entirely, and its error peaks exactly where the flow departs from
horizontal homogeneity, in the channeled acceleration zones.
The classical
and neural field models behave differently: both lose accuracy in the
building-induced regimes relative to open flow (kriging 0.42/0.45 versus 0.33;
UNet 0.42/0.40 versus 0.34), which quantifies wake and channel structure as
the residual difficulty of the benchmark; the neural surrogate matches kriging
in the wake, leads clearly in the channeled regime, and is the most
consistent across cases, while open flow is served equally well by classical
interpolation.
The error map in
panel~(b) makes the mechanism visible: residuals organize along the smoothed
wake boundaries, that is, along shear layers.

\begin{figure}[htbp]
\centering
\includegraphics[width=\linewidth]{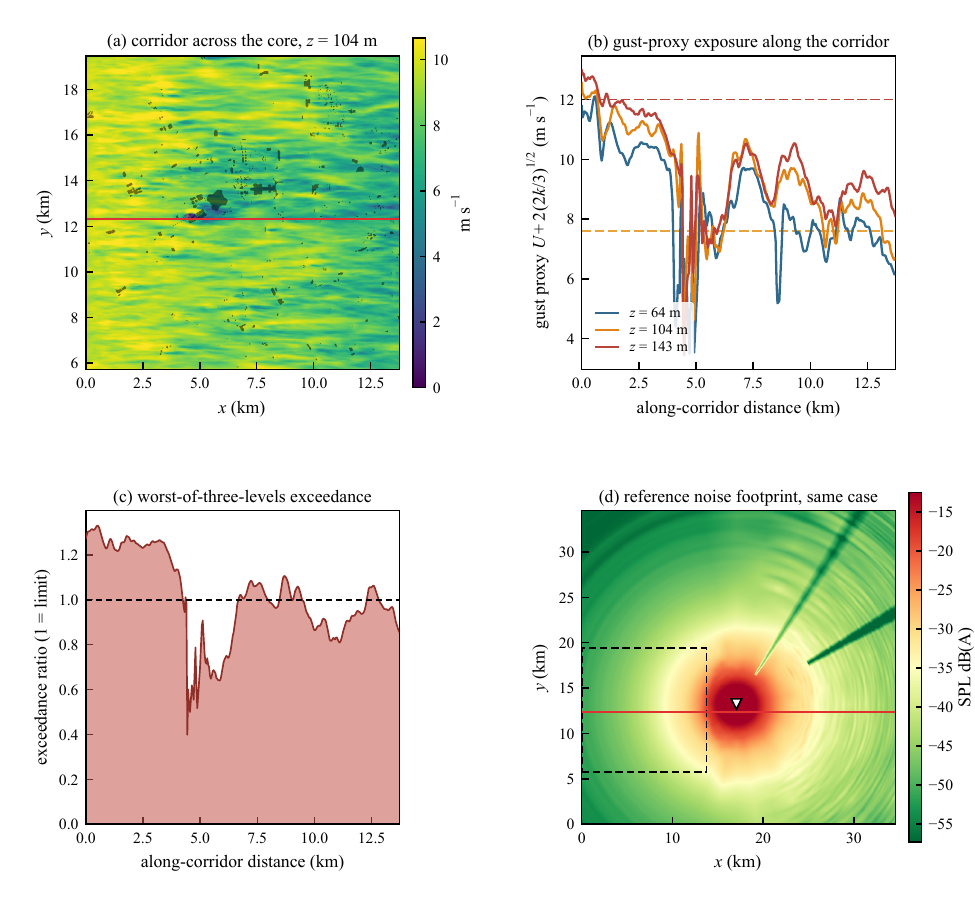}
\caption{Corridor case study (u06 annual, $270^\circ$ inflow).
(a) Transect through the core at $z=104$\,m; (b) gust proxy along the corridor
at three flight levels against the 7.6 and 12\,m/s thresholds; (c)
worst-of-three-levels exceedance ratio (1 = at the limit); (d) reference noise
footprint of the same case with the corridor and the core window marked.}
\label{fig:phys_corridor}
\end{figure}

Figure~\ref{fig:phys_corridor} closes with an operational case study: a
13.7\,km corridor through the core at three flight levels (u06 annual,
$270^\circ$ inflow). The gust proxy $\bar U + 2\sqrt{2k/3}$ drops from about
13\,m/s upstream to below 4\,m/s inside the core wake within roughly one
kilometer, and the worst-of-three-levels exceedance ratio sits above the limit
over most of the transect, collapses to 0.4--0.6 in the wake lull, and
re-crosses the limit repeatedly downstream. Two operational consequences follow.
First, the threshold state of a straight corridor changes on sub-kilometer
scales, so corridor assessment requires fields at the released resolution.
Second, altitude is a genuine degree of freedom: the highest level carries the
largest sustained exposure upstream while the lowest level sees the deepest
wake deficit, so level selection can trade gust margin against
schedule.
The reference noise footprint for the same case shows building shadow wedges
tens of decibels deep cutting the free-field ring pattern, so wind hazard and
community-noise exposure vary along the same corridor and can be screened
jointly from the released fields.

\subsection{Downstream Application Tasks}
Consistent with the task hierarchy of Section~\ref{subsec:formulation}, the full
results for the three downstream application tasks are reported in
\ref{app:task3} (airworthiness wind-compliance risk, including the
path-planner baselines), \ref{app:task4} (site wind-exposure ranking), and
\ref{app:task5} (noise propagation).

\section{Discussion\label{sec:5}}

\subsection{Results Discussion}

Through benchmarking the five core tasks of the \textsc{U3DWind} dataset, this study systematically reveals the underlying trade-offs between statistical accuracy and physical fidelity in urban micro-meteorological surrogate modeling. In the evaluation of wind-field surrogate modeling (Task 1), the experimental results explicitly highlight the prevalence of the ``spectral bias'' phenomenon \citep{Li2021FNO,Lu2021DeepONet}. The reported spectral metric measures the spatial spectral content of the predicted time-averaged fields. The lowest spectral error belongs to 3D kriging, the skip-connected UNet family sits ahead of the neural operators, and the physics-informed variant shows no spectral advantage; preserving spatial spectral content therefore remains an open axis on which no learned baseline dominates, and embedding stronger physical priors into generative architectures remains an open direction for future work. Furthermore, the sensor scaling experiments in Task 2 show that physical priors and learned representations are complementary across sensor densities \citep{Fukami2021Voronoi,Willcox2006Gappy}. At the default sensor density the two prior-based reconstructions lead the benchmark, with Kalman assimilation \citep{Anderson2001EnKF} and Gappy POD essentially tied, while at the sparsest 16-sensor tier the trained sensor transformer generalizes best, so the appropriate reconstruction recipe follows the sensing regime. This finding holds substantial engineering value for low-altitude urban airspace management \citep{Bauranov2021UAMAirspace,Mueller2017UAMAirspace} constrained by physical infrastructure, proving that high-quality offline simulation data can serve as robust prior knowledge to support online, real-time situational awareness.

From the macroscopic perspective of UAM operations and societal acceptance, the results of Tasks 3, 4, and 5 illustrate how the released fields support downstream analyses ranging from aerodynamic risk and site-selection screening to community noise exposure. Once whole inflow slots are withheld, climatology-aware aggregation and pedestrian-comfort heuristics remain competitive with the learned rankers, and they transfer more robustly to the unseen city (\ref{app:task4}). Decision quality for siting therefore traces to the building-resolved exposure statistics a method can access, which every method in the comparison draws from the released fields. In particular, the noise propagation study in Task 5 shows that the urban building canopy, and to a lesser extent the wind field, deform the SPL footprint far from the smooth radial pattern assumed by the free-field screening commonly used in UAM noise assessment. We argue that building-aware acoustic footprint models must become a core component of dynamic flight dispatching for future UAM routes. Ultimately, the measured acceleration of the learned emulators, sub-second per-case inference for the Task 1 surrogates against a full multi-GPU LBM-LES solve, and speedups of $155\times$ to $2178\times$ for the Task 5 acoustic surrogates (Table~\ref{tab:task5}), validates the algorithm-enabling value of the \textsc{U3DWind} dataset and illustrates a viable pathway toward fast surrogate-in-the-loop evaluation for urban wind screening.

\subsection{Limitations and Future Work}
\label{subsec:limits}

However, several limitations remain.

First, the Task 3 wind-compliance label is an operational screening criterion grounded in three published thresholds; a formally promulgated airworthiness wind rule for autonomous urban eVTOL aircraft is still pending. The Task 3 risk score $\hat r$ is the worst exceedance ratio of three thresholds: $7.6$ m/s sustained-wind \citep{Vascik2017Constraint}, the $7.62$ m/s ($25$ ft/s) discrete-gust velocity used as the design-gust criterion in FAR §23.341 / §25.341, and a $12.0$ m/s ground-operations limit drawn from publicly-disclosed eVTOL prototype envelopes. The operational rules eventually promulgated for autonomous urban eVTOL aircraft may evolve as airworthiness standards are formalized. The Task 4 ground truth is the per-case 95th-percentile horizontal wind in the approach cylinder above each candidate rooftop, sampled directly from the LBM-LES output. Tasks 1, 2, and 5 use the LBM-LES output (T1/T2) or a building-diffraction (Maekawa / ISO 9613-2) plus parabolic-equation wind-refraction solver (T5) as ground truth. Validating that the Task 3 worst-of-three exceedance reproduces operator-defined go/no-go decisions in commercial eVTOL flight logs, and that the Task 4 approach-cylinder P95 wind correlates with municipal vertiport-siting decisions, remains downstream work.

Second, the main benchmark remains anchored on Shanghai, and the cross-city evidence covers one held-out city: the core surrogate task and the site-ranking task are additionally evaluated under a train-on-Shanghai, test-on-Hong-Kong hold-out (Section~\ref{subsec:ccgen} and \ref{app:task4}), which quantifies single-city transfer degradation but does not yet bound generalization across all five morphologies. A full leave-one-city-out sweep over the released corpus, together with cross-city transfer learning and foundation-model approaches \citep{Pathak2022FourCastNet}, is future work.

Third, the current dataset primarily focuses on steady or time-averaged flow-field characteristics. While sufficient for preliminary route planning and risk assessment, it cannot capture sudden wind gusts or unsteady vortex street structures at extremely short time scales \citep{Ahmad2017GustLBM}. Considering the extreme sensitivity of UAVs to transient wind disturbances, future dataset iterations should incorporate unsteady Large Eddy Simulation (LES) sequences \citep{Maronga2020PALM} with higher spatio-temporal resolution, alongside explorations into high-fidelity unsteady flow generation techniques based on diffusion models \citep{Du2024DiffusionTurb}. Finally, the volumetric benchmark tasks use downsampled inputs to reduce computational requirements, whereas the complete dataset is publicly available at the original 10~m resolution. Future work will focus on super-resolution enhancement techniques and multi-city morphological co-training to further bridge the gap between simulation data and real-world complex urban operational environments.

\section{Conclusion\label{sec:6}}

This study presents \textsc{U3DWind}, a large-scale building-resolved urban micro-meteorological benchmark dataset and evaluation platform specifically designed for Urban Air Mobility (UAM). By integrating high-resolution Lattice Boltzmann Method with Large-Eddy Simulation (LBM-LES) simulations across 16 inflow directions, three wind speed tiers, three atmospheric scenarios, and five representative Chinese megacities, \textsc{U3DWind} bridges a critical gap in building-resolved low-altitude wind field data. It provides an empirical basis for research on aerodynamic risk assessment and infrastructure planning in the emerging low-altitude economy. The benchmark couples two core wind-field tasks (surrogate prediction and sparse reconstruction) with three downstream application illustrations, providing an operationally oriented assessment suite. Our results demonstrate that deep operator networks preserve the dominant physical structure of the LBM-LES reference field at the urban-canopy scale while reducing per-case inference to sub-second wallclock, thereby establishing the computational feasibility of surrogate-based, city-scale wind screening.

Beyond predictive performance, the underlying scientific findings highlight that the deep integration of physical priors with data-driven architectures is essential for characterizing complex urban flows. Empirical analyses show that with extremely sparse sensing, simple spatial interpolation fails to capture the non-linear flow effects induced by non-uniform urban topologies, while for candidate-site ranking the slot hold-out shows climatology-aware heuristics remaining competitive with learned rankers; in both settings, decision quality traces to how much building-resolved flow structure a method can access. This underscores the necessity of transforming building-resolved offline simulation knowledge into online decision-making intelligence. Furthermore, the findings on noise propagation indicate that building diffraction, with a secondary wind-refraction modulation, deforms the acoustic footprint far from a free-field assumption, so that building-aware ray-tracing is a necessary complement to free-field screening for community noise estimation around vertiports. Looking forward, the open-source release of \textsc{U3DWind} together with our \textsc{LatticeUrbanWind} is intended to serve as a unified experimental testbed for both academia and industry. By fostering exploration into cross-city transfer learning and unsteady flow generation, we hope this dataset and benchmark provide a useful empirical foundation for wind-aware research in urban air mobility. These contributions carry the limitations of Section~\ref{subsec:limits}: the fields are simulated and time-averaged, the benchmark is Shanghai-anchored with a single held-out transfer city, and TKE at the production resolution is a resolved-scale indicator.

\section*{CRediT authorship contribution statement}
\textbf{Shixiong Zhou}: Writing -- original draft, Visualization, Software, Methodology, Data curation, Conceptualization;
\textbf{Huanxia Wei}: Writing -- original draft, Visualization, Software, Resources, Methodology, Data curation, Conceptualization;
\textbf{Chao Xia}: Writing -- review \& editing, Validation, Resources;
\textbf{Yingying Xing}: Writing -- review \& editing, Supervision;
\textbf{Changmin Jiang}: Validation, Methodology;
\textbf{Hai Yang}: Writing -- review \& editing, Supervision;
\textbf{Shuai Jia}: Writing -- review \& editing, Supervision, Project administration, Funding acquisition, Conceptualization.

\section*{Declaration of competing interest}
The authors declare that they have no known competing financial interests or personal relationships that could have appeared to influence the work reported in this paper.

\section*{Acknowledgements}
This work was supported in part by the National Natural Science Foundation of China (Grant No. 72542013) and the Hong Kong Strategic Public Policy Research Funding Scheme (Grant No. S2024.A7.022.24S), awarded to Prof. Shuai Jia. The authors would like to clarify that any information, opinions, findings, conclusions, or recommendations in this paper do not represent the views of the Government of the Hong Kong SAR and/or the SPPRF Project Assessment Panel.

\section*{Data availability}
\dataavailabilitystatement

\section*{Supplementary material}
Supplementary material associated with this article can be found in the online version at \href{https://doi.org/10.1016/j.ast.2026.114092}{\nolinkurl{10.1016/j.ast.2026.114092}}.

\appendix
\counterwithin*{figure}{section}
\counterwithin*{table}{section}
\section{Task 3 results: airworthiness wind-compliance risk}
\label{app:task3}

The experimental results for Task 3, summarized in Table~\ref{tab:task3}, evaluate the models' capacity to regress a continuous aerodynamic risk score $\hat r \in [0, 1]$ along a UAM trajectory. The label is grounded in three operational wind / gust thresholds: each trajectory's risk is the worst exceedance ratio of (i) the $7.6$ m/s sustained-wind operational threshold (the 17 mph UAM operating limit \citep{Vascik2017Constraint}), (ii) a peak gust of $7.62$ m/s ($25$ ft/s discrete gust velocity used as the design gust criterion in FAR §23.341 and §25.341), and (iii) a $12.0$ m/s ground-operations limit drawn from publicly-disclosed eVTOL prototype envelopes such as the EHang 216 and Volocopter VC-series. Because the released fields are time-averaged, the peak gust entering criterion (ii) is a TKE-derived proxy, $U_{\mathrm{gust}} = \lVert \bar{\mathbf{u}}_h \rVert + 2\sqrt{2k/3}$: the mean horizontal speed plus a peak factor of two on the isotropic velocity standard deviation $\sigma_u = \sqrt{2k/3}$. The proxy inherits the TKE resolution caveat of Section~\ref{subsubsec:aijval}. We supervise on $\hat r$ directly and report regression error together with rank correlation; threshold-derived go/no-go metrics ($\text{AUC-ROC}@0.5$, $\text{Brier}@0.5$, expected cost@0.5) use the binary ``operationally-unsafe'' label at $\hat r \ge 0.5$ (pooled positive rate $0.6285$ on the $18{,}000$-trajectory IE test split). Because the IE test split concatenates nine inflow slots whose risk distributions barely overlap, pooled AUC-ROC saturates at $\ge 0.999$ for every learned model and measures slot separation only; we therefore adopt within-inflow-slot macro means, averaged over the six of nine slots that contain both classes, as the primary metrics, and quote the pooled AUC-ROC only as an explicitly labeled secondary column. The analysis reveals three distinct tiers of baseline behavior.

Under the slot-macro metric the physics-based score baselines split into a functioning family and a failing one. The \textit{Gust exceedance count} is a genuine physics baseline at slot-macro AUC-ROC $0.9711$, and the \textit{AIJ-adapted comfort score} reaches $0.9857$; the \textit{Gradient Richardson proxy} follows at $0.9338$, although its calibration is poor (Brier@0.5 $= 0.3222$). The failures are the \textit{GEV peaks-over-threshold} fit ($0.7533$), the \textit{Energy-weighted shear integral} ($0.7102$), and the \textit{Vehicle control-effort proxy} ($0.6602$, the weakest classifier in the table): their scalar scores track inflow strength but not the within-slot flow geometry that decides which trajectories inside a given slot exceed the operational thresholds. The \textit{Constant-risk predictor} emits a degenerate constant score, so its threshold-classification cells are undefined; it serves only as the regression floor at MAE $0.1023$.

Once the slot-separation artifact is removed, shallow learners remain competitive while staying clear of saturation. \textit{Random Forest} reaches slot-macro AUC-ROC $0.9909$ with the lowest expected cost in the table ($0.0290$), \textit{XGBoost} \citep{Chen2016XGBoost} and its \textit{Conformalised} variant \citep{Stankeviciute2021Conformal} reach $0.9912$, and the \textit{Gaussian-process classifier} is a functioning probabilistic baseline at $0.9914$ with expected cost $0.0292$; the AUC-ROC $\ge 0.999$ saturation reported by the pooled view is thus an artifact of slot separation; the within-slot columns carry the discriminative signal. On the regression side the tree boosters retain solid within-slot rank fidelity ($\rho_s = 0.7865$ for Random Forest and $0.7864$ for XGBoost) at MAE $= 0.0158$ and $0.0156$ respectively.

Sequence and field-aware neural architectures hold the top of the classification column and the whole of the regression column. The \textit{GraphSAGE corridor GNN} \citep{Hamilton2017GraphSAGE} and the deep ensemble of Transformers share the best slot-macro AUC-ROC ($0.9925$ for both), with GraphSAGE also posting the lowest Brier@0.5 ($0.0097$); the \textit{Bidirectional LSTM} follows at $0.9919$, and the \textit{Temporal Transformer} \citep{Vaswani2017AttentionIsAllYouNeed} at $0.9904$. Within-slot rank correlation is markedly harder than the pooled view suggested: the best slot-macro Spearman is $0.8215$, attained by the \textit{Heteroscedastic} field-conditioned encoder, which also drives MAE to the table minimum ($0.0122$) at the cost of a higher Brier ($0.0177$); the BiLSTM ($\rho_s = 0.7431$), GraphSAGE ($0.7185$), and the Transformer ensemble ($0.7117$) form the next rank tier, while the dilated \textit{1D-CNN} \citep{Bai2018TCN} trails at $\rho_s = 0.5586$.

\begin{figure}[htbp]
    \centering
    \includegraphics[width=\linewidth]{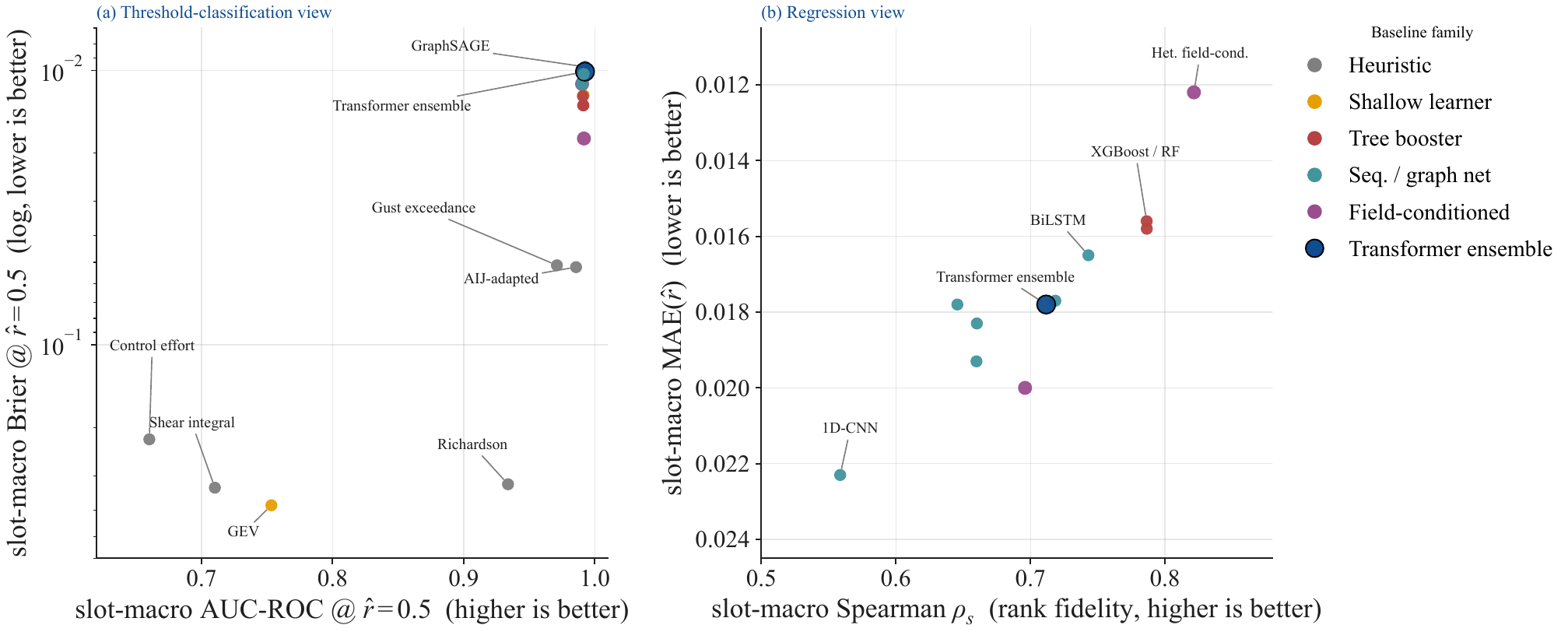}
    \caption{Task 3 baseline landscape on the Shanghai release. Markers are color-coded by baseline family. Panel a is the threshold-classification view: slot-macro AUC-ROC@$\hat r=0.5$ on the horizontal axis, slot-macro Brier@$\hat r=0.5$ on the vertical (log scale, axis inverted, so up = better). The GraphSAGE corridor GNN and the Transformer ensemble lead at AUC-ROC $= 0.9925$, with GraphSAGE attaining the lowest Brier in the table ($0.0097$); the GEV, shear-integral, and control-effort scores collapse toward the low-AUC corner (AUC-ROC $\le 0.7533$). Panel b is the regression view: slot-macro Spearman $\rho_s$ on the horizontal, slot-macro MAE($\hat r$) on the vertical (axis inverted). The heteroscedastic field-conditioned encoder occupies the best corner ($\rho_s = 0.8215$, MAE $= 0.0122$); Random forest, XGBoost, the BiLSTM, GraphSAGE, and the Transformer ensemble follow at $\rho_s \in [0.71, 0.79]$, while the dilated 1D-CNN trails at $\rho_s = 0.5586$.}
    \label{fig:task3_risk}
\end{figure}

Operationally, the conclusion is that the operational-threshold-grounded Task 3 label is non-trivial once the slot-separation artifact is removed: pooled AUC-ROC saturates at $\ge 0.999$ for every learned model, whereas the within-slot macro view separates the field, with graph- and ensemble-based models at slot-macro AUC-ROC $0.9925$, the Gaussian-process classifier and the tree boosters within $0.002$ of that lead, functioning physics baselines near $0.93$ to $0.99$, and the GEV, shear-integral, and control-effort scores failing below $0.76$. The residual split between the best classifiers (GraphSAGE and the Transformer ensemble) and the best regressor (the heteroscedastic field-conditioned encoder, $\rho_s = 0.8215$, MAE $= 0.0122$) is the natural trade-off between binary go/no-go calibration at the operational threshold and the continuous risk gradient that supports dynamic route re-planning. Fig.~\ref{fig:task3_risk} renders the same baseline set from two complementary empirical views: panel a places every baseline on (slot-macro AUC-ROC@0.5, slot-macro Brier@0.5) so the threshold-classification tier is visible, and panel b places the regression-capable baselines on (slot-macro Spearman $\rho_s$, slot-macro MAE) so the regression tier is visible.

\begin{table}[t]
\centering
\caption{Task 3 risk-scoring baselines on the Shanghai \textsc{U3DWind} release (20 baseline models). The continuous risk label $\hat r$ is the worst (largest exceedance ratio) of three operational wind thresholds: the $7.6$ m/s sustained-wind operational threshold \citep{Vascik2017Constraint}, the $7.62$ m/s ($25$ ft/s) peak gust used as the design gust criterion in FAR §23.341 / §25.341, and a $12.0$ m/s ground-operations limit matching publicly-disclosed eVTOL prototype envelopes. The exceedance ratio is rescaled so that $\hat r = 0.5$ corresponds to $70$\,\% of the operational threshold and $\hat r = 1$ saturates at $1.4$x the threshold. Threshold-derived metrics (AUC-ROC@0.5, Brier@0.5, Exp.\,cost@0.5) use the binary label ``operationally-unsafe at $\hat r \ge 0.5$'' (pooled positive rate $0.6285$ on the $18{,}000$-trajectory IE test split). Because the IE test split concatenates nine inflow slots whose risk distributions barely overlap, pooled scores saturate for every learned model; the primary columns are therefore within-inflow-slot macro means over the six of nine inflow slots that contain both classes, and the pooled AUC-ROC@0.5 is quoted only as an explicitly labeled secondary column. Neural baselines report mean $\pm$ std over three seeds; all other rows are single deterministic runs. Dashes mark undefined cells: the constant-risk baseline emits a degenerate constant score, so its threshold-classification cells are undefined, and score-only baselines produce no calibrated regression output. Best value per column is bold.}
\label{tab:task3}
\footnotesize
\setlength{\tabcolsep}{2pt}
\begin{tabular}{lcccccc}
\toprule
 & \multicolumn{5}{c}{Within-inflow-slot macro means (primary)} & Pooled (secondary) \\
\cmidrule(lr){2-6}\cmidrule(lr){7-7}
Baseline & AUC-ROC@0.5 $\uparrow$ & Brier@0.5 $\downarrow$ & Exp.\,cost@0.5 $\downarrow$ & Spearman $\rho_s$ $\uparrow$ & MAE($\hat r$) $\downarrow$ & AUC-ROC@0.5 $\uparrow$ \\
\midrule
Gust exceedance count             & $0.9711$ & $0.0513$ & $0.0369$ & --- & --- & $0.9970$ \\
AIJ-adapted comfort score         & $0.9857$ & $0.0521$ & $0.2171$ & --- & --- & $0.9964$ \\
Gradient Richardson proxy         & $0.9338$ & $0.3222$ & $0.3836$ & --- & --- & $0.8582$ \\
GEV peaks-over-threshold          & $0.7533$ & $0.3843$ & $4.0111$ & --- & --- & $0.7630$ \\
Energy-weighted shear integral    & $0.7102$ & $0.3316$ & $0.3685$ & --- & --- & $0.7068$ \\
Vehicle control-effort proxy      & $0.6602$ & $0.2210$ & $3.6999$ & --- & --- & $0.9162$ \\
Constant-risk predictor           & --- & --- & --- & --- & $0.1023$ & --- \\
\midrule
Random forest                     & $0.9909$ & $0.0124$ & $\mathbf{0.0290}$ & $0.7865$ & $0.0158$ & $0.9991$ \\
XGBoost                           & $0.9912$ & $0.0134$ & $0.0315$ & $0.7864$ & $0.0156$ & $0.9991$ \\
XGBoost (conformalised)           & $0.9912$ & $0.0134$ & $0.0315$ & --- & --- & $0.9991$ \\
Gaussian-process classifier       & $0.9914$ & $0.0123$ & $0.0292$ & --- & --- & $0.9992$ \\
\midrule
Dilated 1D-CNN                    & $0.9901 \pm 0.0008$ & $0.0111 \pm 0.0001$ & $0.0675 \pm 0.0166$ & $0.5586 \pm 0.0419$ & $0.0223 \pm 0.0021$ & $0.9990$ \\
Bidirectional LSTM                & $0.9919 \pm 0.0004$ & $0.0103 \pm 0.0002$ & $0.0517 \pm 0.0068$ & $0.7431 \pm 0.0123$ & $0.0165 \pm 0.0017$ & $0.9992$ \\
Temporal Transformer              & $0.9904 \pm 0.0008$ & $0.0110 \pm 0.0003$ & $0.0686 \pm 0.0134$ & $0.6602 \pm 0.0568$ & $0.0183 \pm 0.0002$ & $0.9991$ \\
Physics-guided Transformer        & $0.9904 \pm 0.0010$ & $0.0113 \pm 0.0005$ & $0.0531 \pm 0.0141$ & $0.6457 \pm 0.0231$ & $0.0178 \pm 0.0001$ & $0.9991$ \\
GraphSAGE corridor GNN            & $\mathbf{0.9925} \pm 0.0004$ & $\mathbf{0.0097} \pm 0.0001$ & $0.0630 \pm 0.0009$ & $0.7185 \pm 0.0080$ & $0.0177 \pm 0.0007$ & $\mathbf{0.9993}$ \\
Deep ensemble of Transformers     & $\mathbf{0.9925} \pm 0.0002$ & $0.0101 \pm 0.0002$ & $0.0678 \pm 0.0065$ & $0.7117 \pm 0.0203$ & $0.0178 \pm 0.0008$ & $\mathbf{0.9993}$ \\
MC-dropout Transformer            & $0.9915 \pm 0.0009$ & $0.0103 \pm 0.0005$ & $0.0520 \pm 0.0035$ & $0.6599 \pm 0.0568$ & $0.0193 \pm 0.0022$ & $0.9992$ \\
Field-conditioned encoder         & $0.9903 \pm 0.0017$ & $0.0112 \pm 0.0016$ & $0.0735 \pm 0.0348$ & $0.6960 \pm 0.0511$ & $0.0200 \pm 0.0052$ & $0.9990$ \\
Heteroscedastic field-conditioned & $0.9917 \pm 0.0004$ & $0.0177 \pm 0.0010$ & $0.0636 \pm 0.0026$ & $\mathbf{0.8215} \pm 0.0046$ & $\mathbf{0.0122} \pm 0.0000$ & $0.9991$ \\
\bottomrule
\end{tabular}
\end{table}

Table~\ref{tab:task3_inputs} lists the exact inputs each risk baseline
consumes. The Task 3 label is a deterministic functional of the along-path
environment (the worst exceedance ratio of the horizontal wind plus the TKE
gust term against the three operational thresholds); every baseline receives
raw, normalized wind and TKE samples along the trajectory, and no baseline
receives the functional's output, the thresholds, or any exceedance ratio.
The gust-exceedance row deliberately re-evaluates the label functional on the
20-waypoint sampled track and is marked as the physics reference; its gap to
the learned sequence models (slot-macro AUC $0.9711$ against $0.9925$)
measures what full-track integration adds over the sampled-track functional,
and is the non-trivial headroom of the task.

\begin{table}[htbp]
\centering
\caption{Inputs consumed by each Task 3 risk baseline. No baseline receives
the label functional, the operational thresholds, or any exceedance ratio;
``derived shear'' is computed from the wind samples inside the model input
pipeline.}
\label{tab:task3_inputs}
\footnotesize
\setlength{\tabcolsep}{4pt}
\begin{tabular}{p{0.28\textwidth} p{0.62\textwidth}}
\toprule
Baseline & Inputs \\
\midrule
Gust exceedance & analytic worst-exceedance functional on the sampled track (horizontal wind + TKE gust term); physics reference for the label functional \\
Shear integral & along-path wind-increment integral $\sum |\Delta U|/\Delta s$ \\
Richardson & bulk Richardson number from along-path wind shear with an assumed stable $N^2$ \\
AIJ-adapted & along-path wind-speed exceedance frequency (comfort-criterion form) \\
Control effort & trajectory waypoints + along-path wind increments \\
GEV tail & generalized-extreme-value fit of along-path wind maxima \\
Constant risk & none (degenerate reference) \\
\midrule
Random forest / XGBoost / conformal XGBoost & 16 summary statistics: mean/std/max of horizontal wind, vertical wind, TKE, derived shear; path length; mean altitude; shear- and wind-peak positions \\
GP classifier & 6 statistics: wind-level mean and max, fluctuation std and p95, TKE mean and max \\
\midrule
1D-CNN / BiLSTM / Transformer / GraphSAGE / PI-Transformer & per-waypoint $(20 \times 8)$ tensor: waypoint $xyz$, wind $uvw$, TKE, derived shear \\
Transformer ensemble / MC-dropout & same per-waypoint tensor (wrapper variants of the Transformer backbone) \\
Field-conditioned encoder (+ heteroscedastic) & per-waypoint tensor + pooled 3D wind-field tokens from the case volume \\
\bottomrule
\end{tabular}
\end{table}

Path-planner baselines on the same Task 3 trajectory pool are summarized separately in Table~\ref{tab:task3_planners}; they consume the risk score as an obstacle map and produce 4D corridors, so the natural metrics are corridor length, total energy, mission feasibility, regret against the MILP optimum, and wall-clock time. Among the classical search planners, \textit{Dijkstra} attains the highest feasibility ($0.990$) at the lowest wall-clock cost ($25.1$ s); \textit{D-star Lite} delivers a marginally lower energy than the wind-aware A-star variant at roughly three times its wall-clock cost. The dynamic-programming \textit{DP} planner finds the shortest corridors (length $3376$ m) but at a $51{,}300$ s wall-clock cost. \textit{iLQR} and \textit{MPC} give a different trade-off: they find more energy-efficient corridors than MILP (best energy $75{,}666$ J for iLQR) but with sharply lower feasibility ($0.560$ and $0.480$ respectively) because their continuous-time dynamics model has no formal collision-avoidance guarantee on the discretised obstacle map. Optimisation-under-uncertainty planners are a third tier: \textit{Chance-constrained MILP} stays close to \textit{MILP} on length ($3493$ versus $3524$ m) and feasibility ($0.970$ versus $0.980$), while \textit{Risk-averse MPC} trades feasibility for robustness. Among learning-based planners, \textit{Behavioral Cloning} (\textit{bc}) and \textit{PPO} fail to match the classical baselines on length and feasibility, while the trajectory-token \textit{Transformer} (\textit{traj\_transformer}) reaches a usable feasibility of $0.750$ at moderate length inflation, and the \textit{GNN-Planner} attains the highest learning-based feasibility ($0.930$) at the longest corridors in the table ($6293$ m).

\begin{table}[htbp]
\centering
\caption{Task 3 path-planner baselines on the Shanghai \textsc{U3DWind} release. Length is the corridor arc-length in meters, energy is the integrated power along the corridor, feasibility is the fraction of cases where the planner returned a collision-free corridor, regret-vs-MILP compares the corridor cost against the MILP optimum (negative is better), and wall is the planner's wall-clock time. Best value per column is marked bold. The planner rows are single deterministic runs retained from the initial evaluation and are excluded from the automated regeneration described in \ref{app:protocol}.}
\label{tab:task3_planners}
\footnotesize
\begin{tabular}{lccccc}
\toprule
Baseline & Length (m) $\downarrow$ & Energy $\downarrow$ & Feas. $\uparrow$ & Regret vs MILP (\%) $\downarrow$ & Wall (s) $\downarrow$ \\
\midrule
Dijkstra                  & $3513$           & $79215$           & $\mathbf{0.990}$ & n/a               & $\mathbf{25.1}$    \\
A-star                    & $3558$           & $80194$           & $0.980$          & n/a               & $71.5$             \\
A-star with wind          & $3502$           & $78542$           & $0.960$          & $-0.341$          & $11050.1$          \\
D-star Lite               & $3463$           & $77729$           & $0.980$          & $-1.365$          & $34569.0$          \\
DP planner                & $\mathbf{3376}$  & $76108$           & $0.980$          & n/a               & $51299.9$          \\
MILP                      & $3524$           & $79158$           & $0.980$          & $0.000$           & $674.2$            \\
Chance-constrained MILP   & $3493$           & $78390$           & $0.970$          & n/a               & $19311.4$          \\
Worst-case-robust         & $3502$           & $78545$           & $0.980$          & n/a               & $22436.7$          \\
iLQR                      & $3383$           & $\mathbf{75666}$  & $0.560$          & $\mathbf{-4.114}$ & $710.4$            \\
MPC                       & $3434$           & $76484$           & $0.480$          & $-3.058$          & $139.3$            \\
Risk-averse MPC           & $3452$           & $76919$           & $0.640$          & n/a               & $337.3$            \\
Behavioral cloning       & $5881$           & $132153$          & $0.530$          & $89.838$          & $25.2$             \\
PPO                       & $3779$           & $84743$           & $0.000$          & $9.732$           & $318.8$            \\
Trajectory Transformer    & $3962$           & $89294$           & $0.750$          & n/a               & $37.7$             \\
GNN planner               & $6293$           & $141717$          & $0.930$          & n/a               & $51.6$             \\
\bottomrule
\end{tabular}%
\end{table}

\section{Task 4 results: site wind-exposure ranking}
\label{app:task4}

The quantitative evaluation of site wind-exposure ranking, as summarized in Table~\ref{tab:task4}, underscores the difficulty of recovering a volumetric LBM-LES observable (the per-case 95th-percentile horizontal wind speed inside each candidate rooftop's approach cylinder) from noisy point-sample features \citep{Rohrmeier2025Vertiport,Lee2025VertiportLocation}. Under the slot hold-out, in which whole (city, inflow slot) groups are withheld from training, the field compresses into a narrow band: all 21 baselines fall between NDCG@5 $= 0.784$ and $0.829$, with no sharp tier structure.

\begin{table}[t]
\centering
\caption{Task 4 site wind-exposure ranking baselines on the Shanghai \textsc{U3DWind}
release under the SO slot hold-out (whole (city, inflow-slot) groups held out;
64 test queries). Three seeds exist for every model, but the SO seed redraws the
hold-out groups themselves, so cross-seed spread measures split difficulty; by design the table reports the seed-0 protocol only. Kendall
$\tau$ is computed over the 48 of 64 seed-0 test queries with a defined ordering.
Best value per column is bold.}
\label{tab:task4}
\footnotesize
\begin{tabular}{lccccc}
\toprule
Baseline & NDCG@5 & NDCG@10 & Kendall $\tau$ & MRR & P@3 \\
\midrule
ListNet & $\mathbf{0.8291}$ & $0.8089$ & $0.2889$ & $0.6243$ & $0.4896$ \\
DeepSets & $0.8195$ & $\mathbf{0.8121}$ & $0.2973$ & $\mathbf{0.6591}$ & $0.5208$ \\
Windrose-averaged & $0.8191$ & $0.8038$ & $0.2740$ & $0.5560$ & $0.4375$ \\
Borda ensemble & $0.8184$ & $0.8092$ & $0.3003$ & $0.5974$ & $0.4635$ \\
ListMLE & $0.8152$ & $0.7949$ & $0.2892$ & $0.6400$ & $0.5104$ \\
RankNet & $0.8126$ & $0.8032$ & $0.2913$ & $0.6307$ & $0.5000$ \\
GAT-Ranker & $0.8087$ & $0.8076$ & $\mathbf{0.3202}$ & $0.5893$ & $0.4219$ \\
Linear probe & $0.8062$ & $0.7941$ & $0.2687$ & $0.5332$ & $0.4062$ \\
Set Transformer & $0.8060$ & $0.8017$ & $0.2899$ & $0.5515$ & $0.4531$ \\
AHP & $0.8011$ & $0.7947$ & $0.2817$ & $0.5469$ & $0.4531$ \\
Conformal ranking & $0.7994$ & $0.7842$ & $0.2909$ & $0.6431$ & $\mathbf{0.5260}$ \\
VIKOR & $0.7992$ & $0.7847$ & $0.2095$ & $0.5120$ & $0.4323$ \\
LambdaMART & $0.7982$ & $0.7859$ & $0.2587$ & $0.6092$ & $0.4531$ \\
PROMETHEE & $0.7982$ & $0.7919$ & $0.2193$ & $0.5251$ & $0.4219$ \\
IEC availability & $0.7968$ & $0.7784$ & $0.2847$ & $0.6127$ & $0.5000$ \\
NEN-8100 & $0.7953$ & $0.7725$ & $0.2810$ & $0.6020$ & $0.4635$ \\
Bayesian ListMLE & $0.7946$ & $0.7899$ & $0.2956$ & $0.5759$ & $0.4427$ \\
AIJ-adapted & $0.7920$ & $0.7854$ & $0.2787$ & $0.5565$ & $0.4375$ \\
TOPSIS & $0.7893$ & $0.7834$ & $0.2215$ & $0.5164$ & $0.4115$ \\
Lawson & $0.7890$ & $0.7832$ & $0.2865$ & $0.6022$ & $0.4531$ \\
MILP siting & $0.7839$ & $0.7606$ & $0.2325$ & $0.5226$ & $0.4323$ \\
\bottomrule
\end{tabular}
\end{table}

The compression is most visible at the bottom of the old hierarchy. The pedestrian-comfort heuristics (\textit{Lawson}, \textit{NEN-8100}) sit inside the main field: Lawson reaches NDCG@5 $= 0.789$ and NEN-8100 reaches $0.795$, above the table minimum set by \textit{MILP siting} ($0.784$), and their Kendall $\tau$ values ($0.287$ and $0.281$) sit inside the range spanned by the learned rankers. The classical MCDA aggregators split into two behaviors. The outranking and distance-based methods (\textit{TOPSIS}, \textit{VIKOR}, \textit{PROMETHEE}) hold NDCG@5 between $0.789$ and $0.799$ but post the weakest rank correlations in the table (Kendall $\tau = 0.222$, $0.210$, and $0.219$, respectively). The \textit{Windrose-averaged} aggregator, by contrast, is the strongest untrained baseline at NDCG@5 $= 0.819$, third overall and within $0.010$ of the best learned ranker; its climatological weighting transfers to unseen inflow slots nearly as well as trained models do.

The learned rankers still occupy the top of the table, but by a narrow margin. \textit{ListNet} leads at NDCG@5 $= 0.829$, followed by \textit{DeepSets} \citep{Zaheer2017DeepSets} at $0.820$, which also posts the best NDCG@10 ($0.812$) and the best MRR ($0.659$), and the \textit{Borda ensemble} at $0.818$; \textit{ListMLE} ($0.815$) and \textit{RankNet} ($0.813$) complete the leading group. \textit{GAT-Ranker} recovers the highest Kendall $\tau$ in the table ($0.320$) and \textit{Conformal ranking} the highest Precision@3 ($0.526$). Two learned baselines fall back into the untrained band: \textit{LambdaMART} \citep{Burges2010LambdaMART,Chen2016XGBoost} drops to NDCG@5 $= 0.798$, and the \textit{Set Transformer} \citep{Lee2019SetTransformer} at $0.806$ does no better than the \textit{Linear probe} at the same $0.806$. Across all 21 baselines the Kendall $\tau$ range is $0.210$ to $0.320$; no method recovers a strong per-case ordering on held-out inflow slots.

\begin{figure}[htbp]
    \centering
    \includegraphics[width=\linewidth]{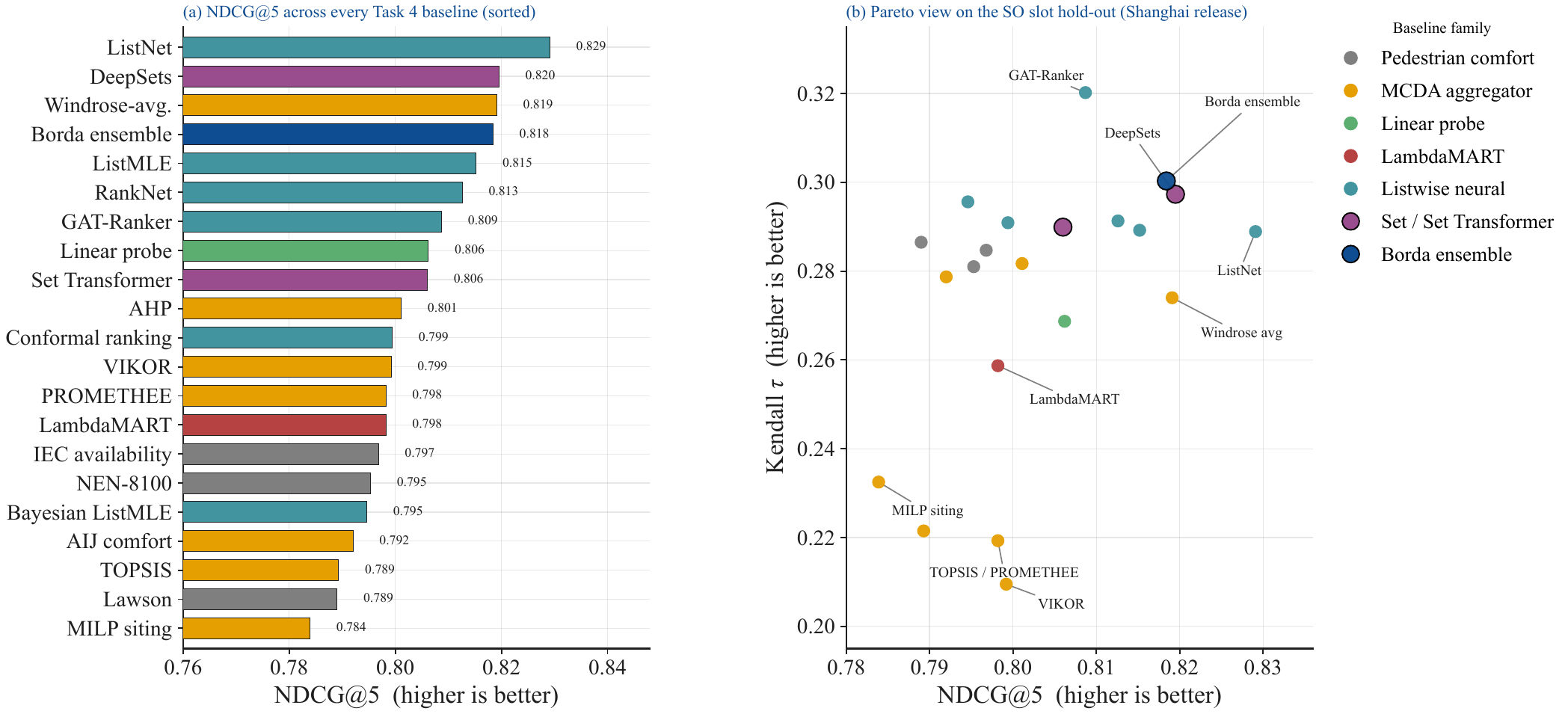}
    \caption{Task 4 baseline landscape under the SO slot hold-out (seed 0, Shanghai release). Both panels read directly from the same per-model metrics that feed Table~\ref{tab:task4}; markers and bars are color-coded by baseline family. Panel a is a horizontal bar chart of every baseline sorted by NDCG@5; the field spans a narrow band from MILP siting ($0.784$) to ListNet ($0.829$), with the untrained Windrose-averaged aggregator ($0.819$) interleaved among the learned rankers. Panel b is the Pareto view on (NDCG@5, Kendall $\tau$): GAT-Ranker attains the highest $\tau$ ($0.320$) and ListNet the highest NDCG@5, while the outranking MCDA methods (TOPSIS, VIKOR, PROMETHEE) occupy the low-$\tau$ region ($\tau \le 0.222$).}
    \label{fig:task4_ranking}
\end{figure}

Learned listwise rankers hold three of the four best NDCG@5 scores, yet the margin of the best learned model over the best untrained aggregator is $0.010$ (ListNet at $0.829$ versus Windrose-averaged at $0.819$), and DeepSets ($0.820$) is separated from Windrose-averaged by less than $0.001$. With a single seed per model, this gap cannot be distinguished from run-to-run variation: learned rankers sit at the top of the table, slightly ahead of classical climatology-aware aggregation, without a decisive advantage once whole inflow slots are withheld from training. Fig.~\ref{fig:task4_ranking} renders the same baseline set from two complementary empirical views: panel a sorts every baseline by NDCG@5, showing a compressed band of scores, and panel b places the same baselines on (NDCG@5, Kendall $\tau$) so the trade-off between the two ranking criteria is visible.

Table~\ref{tab:task4_cc} reports the same baseline set under the cross-city
hold-out (trained on the Shanghai corpus, tested on all 144 Hong Kong queries;
learned rankers averaged over three seeds, deterministic baselines single-run).
The ordering inverts relative to the slot hold-out: the MCDA aggregators transfer robustly and lead the table (VIKOR NDCG@5 $= 0.912$), the linear probe on frozen features holds $0.905$, while the best multi-seed learned ranker (LambdaMART, $0.864$) gives up its within-city advantage, and the set-input rankers additionally carry one collapsed training seed each, which their flat validation loss cannot detect. Kendall $\tau$ stays low throughout, spanning $0.095$ (GAT-Ranker) to $0.231$, a compression that reflects the many near-tied Hong Kong sites.

\begin{table}[htbp]
\centering
\caption{Task 4 site wind-exposure ranking under the cross-city hold-out
(train on Shanghai, test on all 144 Hong Kong queries). Learned rankers report
the mean over three seeds; deterministic baselines are single-run. Best value
per column is bold.}
\label{tab:task4_cc}
\footnotesize
\begin{tabular}{lcccccc}
\toprule
Baseline & Seeds & NDCG@5 & NDCG@10 & Kendall $\tau$ & MRR & P@3 \\
\midrule
VIKOR & 1 & $\mathbf{0.912}$ & $\mathbf{0.899}$ & $0.220$ & $0.837$ & $0.701$ \\
PROMETHEE & 1 & $0.906$ & $0.897$ & $0.217$ & $0.819$ & $0.690$ \\
TOPSIS & 1 & $0.905$ & $0.896$ & $0.218$ & $\mathbf{0.849}$ & $0.704$ \\
Linear probe & 1 & $0.905$ & $0.895$ & $0.216$ & $0.843$ & $\mathbf{0.711}$ \\
Windrose-averaged & 1 & $0.899$ & $0.894$ & $0.218$ & $0.802$ & $0.678$ \\
NEN-8100 & 1 & $0.895$ & $0.878$ & $0.229$ & $0.802$ & $0.688$ \\
AHP & 1 & $0.894$ & $0.891$ & $0.224$ & $0.767$ & $0.692$ \\
MILP siting & 1 & $0.892$ & $0.878$ & $0.222$ & $0.804$ & $0.664$ \\
Lawson & 1 & $0.888$ & $0.887$ & $0.231$ & $0.791$ & $0.671$ \\
AIJ-adapted & 1 & $0.885$ & $0.881$ & $0.224$ & $0.803$ & $0.664$ \\
IEC availability & 1 & $0.882$ & $0.874$ & $\mathbf{0.231}$ & $0.804$ & $0.671$ \\
LambdaMART & 3 & $0.864$ & $0.857$ & $0.191$ & $0.785$ & $0.652$ \\
Borda ensemble & 1 & $0.862$ & $0.854$ & $0.222$ & $0.763$ & $0.625$ \\
ListNet & 3 & $0.855$ & $0.853$ & $0.221$ & $0.744$ & $0.606$ \\
RankNet & 3 & $0.848$ & $0.845$ & $0.216$ & $0.720$ & $0.593$ \\
GAT-Ranker & 3 & $0.814$ & $0.809$ & $0.095$ & $0.613$ & $0.512$ \\
Conformal ranking & 3 & $0.811$ & $0.814$ & $0.200$ & $0.647$ & $0.522$ \\
Set Transformer & 3 & $0.809$ & $0.813$ & $0.169$ & $0.660$ & $0.531$ \\
ListMLE & 3 & $0.808$ & $0.809$ & $0.200$ & $0.630$ & $0.518$ \\
DeepSets & 3 & $0.807$ & $0.808$ & $0.151$ & $0.657$ & $0.540$ \\
Bayesian ListMLE & 3 & $0.801$ & $0.801$ & $0.196$ & $0.622$ & $0.499$ \\
\bottomrule
\end{tabular}
\end{table}

\section{Task 5 results: noise propagation}
\label{app:task5}

\begin{table}[t]
\centering
\caption{Task 5 UAV noise propagation baselines on the Shanghai \textsc{U3DWind} release, scored against the building-diffraction plus wind reference on the IE split (held-out inflow directions 90/270\,deg, directional interpolation; 752 test items). SPL errors are A-weighted dB(A). Wind RMSE is the RMSE of the predicted wind modulation against the reference wind term; baselines with no wind input predict zero modulation, so their Wind RMSE equals the RMS magnitude of the reference wind term, $4.040$\,dB(A), by construction. Shadow IoU is the overlap of the predicted and reference building-shadow masks and Deep IoU restricts the comparison to the deep-shadow cores; the reference shadow covers a fraction $0.163$ of receiver cells. Speedup is the reference label-generation time divided by the per-case inference time. The best learned or surrogate value per column is bold.}
\label{tab:task5}
\footnotesize
\setlength{\tabcolsep}{4pt}
\begin{tabular}{lccccccc}
\toprule
Baseline & $\varepsilon_{L_2}^{\mathrm{SPL}}\downarrow$ (dB) & $\Delta L_{p,A}^{\max}\downarrow$ (dB) & Wind RMSE\,$\downarrow$ (dB) & Shadow IoU\,$\uparrow$ & Deep IoU\,$\uparrow$ & Time (s) & Speedup\,$\uparrow$ \\
\midrule
ISO 9613 free field (no buildings)      & $4.645$ & $26.130$ & $4.040$ & $0.000$ & $0.000$ & $1.105$   & $441.3\times$ \\
Still-air ray (binary occlusion)$^{c}$  & $4.564$ & $25.750$ & $4.040$ & $0.052$ & $0.000$ & $298.968$ & $1.6\times$ \\
Wind-blind MLP per cell                 & $4.007$ & $24.209$ & $4.040$ & $0.032$ & $0.008$ & $0.314$   & $1555.1\times$ \\
Wind-blind 2D-UNet                      & $4.130$ & $23.721$ & $4.040$ & $0.000$ & $0.000$ & $\mathbf{0.224}$ & $\mathbf{2177.6\times}$ \\
Wind-aware 3D-UNet                      & $3.968$ & $23.424$ & $3.211$ & $0.197$ & $0.015$ & $0.702$   & $694.8\times$ \\
Wind-aware FNO                          & $3.051$ & $18.612$ & $3.433$ & $0.328$ & $0.492$ & $1.050$   & $464.5\times$ \\
Diffraction-aware ray surrogate$^{b}$   & $\mathbf{1.821}$ & $\mathbf{8.337}$ & $\mathbf{1.821}$ & $\mathbf{0.397}$ & $\mathbf{0.749}$ & $3.145$ & $155.0\times$ \\
Reference (diffraction $+$ wind)$^{a}$  & --- & --- & --- & --- & --- & $487.6$ & --- \\
\bottomrule
\end{tabular}\\[2pt]
\parbox{\linewidth}{\footnotesize
$^{a}$\,The reference row restates the label, so its error columns are undefined; the quoted $487.6$\,s is the label-generation wallclock per case, dominated by the Crank-Nicolson parabolic-equation wind march ($484.1$\,s), while the still-air diffraction split alone takes $3.5$\,s.
$^{b}$\,The surrogate solves the still-air diffraction field inside its timed prediction path and learns only the wind term, so its SPL error equals its Wind RMSE by construction.
$^{c}$\,The still-air ray baseline's cost is its own line-of-sight occlusion sweep, the dominant term of this baseline by design.}
\end{table}

The results for Task 5, presented in Table~\ref{tab:task5}, quantify how the urban building canopy and the resolved wind field together shape the community noise footprint of a cruising UAV \citep{Cetin2022UAMAcceptance}, and evaluate how well baseline models reproduce it. The scenario is a UAV cruising $120$\,m above the acoustic ground whose noise reaches receivers on a $40$\,m plane: the reference solver propagates each octave band through a Maekawa barrier-diffraction model over the LBM-LES building heights and adds a Crank-Nicolson parabolic-equation (PE) wind-refraction term computed from the same flow field; the full parameter set is given in Table~\ref{tab:task5_params} and the geometry in Fig.~\ref{fig:task5_schematic}. The building canopy is the dominant effect, casting deep, spatially structured acoustic shadows, while the wind imposes a secondary directional modulation of a few dB(A). Conventional free-field screening that ignores the buildings is therefore insufficient for low-altitude UAM operations \citep{Pascioni2018UAVNoise,Tan2021UAMNoise}.

Table~\ref{tab:task5_params} specifies the reference acoustic model in full.
The source is a monopole with a broadband A-weighted level of $L_W = 80$~dB(A)
(electric-quadrotor scale), split equally across nine octave bands (31.5~Hz to
8~kHz), with a cardioid directivity peaking along the downward thrust axis
(floor $-13$~dB). Propagation combines spherical free-field spreading
($-20\log_{10} r - 11$), ISO~9613-1 atmospheric absorption at $20^\circ$C,
70\,\% relative humidity, and $101.325$~kPa, per-band Maekawa/Fresnel barrier
diffraction over the building-top height profile sampled along 144 azimuth
rays, and a parabolic-equation wind-refraction term marched with an azimuthal
effective sound speed derived from the simulated $(u, v)$ field. The PE march
covers the five bands from 31.5 to 500~Hz; higher bands carry a zero wind correction, and the per-band coverage
is recorded in provenance keys that travel with every released run. No ground
reflection image and no turbulence-scattering model are included.

The source model is deliberately screening-level: it carries no rotor tonal
content or blade-passing harmonics, no load-dependent broadband spectrum, and
no motion or Doppler effects; each footprint is computed for a static source
position on the evaluation grid. The task therefore benchmarks community-scale exposure patterns shaped by buildings and wind; vehicle-specific aeroacoustics lies outside its scope, and coupling measured eVTOL source spectra and moving sources into
the same propagation chain is future work.

\begin{table}[htbp]
\centering
\caption{Reference acoustic model of Task 5, as implemented in the released
label-generation code.}
\label{tab:task5_params}
\footnotesize
\begin{tabular}{ll}
\toprule
Source & monopole, $L_W = 80$~dB(A) broadband, equal split over 9 octave bands (31.5~Hz--8~kHz) \\
Directivity & cardioid, peak along downward thrust axis, floor $-13$~dB \\
Geometry & source $120$~m and receiver plane $40$~m above the acoustic ground (terrain top) \\
Spreading & spherical free field, $-20\log_{10} r - 11$; no ground-reflection image \\
Atmospheric absorption & ISO~9613-1 at $20^\circ$C, 70\,\% RH, $101.325$~kPa \\
Building diffraction & Maekawa/Fresnel per band, building-top profiles along 144 azimuth rays \\
Wind refraction & Crank--Nicolson PE, azimuthal effective sound speed from $(u,v)$; bands 31.5--500~Hz \\
Bands above 500~Hz & zero wind correction; coverage recorded in provenance keys \\
Turbulence scattering & not modeled \\
\bottomrule
\end{tabular}
\end{table}

\begin{figure}[htbp]
\centering
\begin{tikzpicture}[x=0.055cm, y=0.028cm, font=\small]
% ground and buildings
\draw[thick] (0,0) -- (200,0);
\fill[gray!35] (18,0) rectangle (30,55);
\fill[gray!35] (44,0) rectangle (58,90);
\fill[gray!35] (74,0) rectangle (86,38);
\fill[gray!35] (104,0) rectangle (120,70);
\fill[gray!35] (140,0) rectangle (152,48);
\node[gray!60!black] at (97,-14) {buildings (LBM-LES height map)};
% receiver plane
\draw[dashed, blue!60!black] (0,40) -- (200,40) node[right=2pt, blue!60!black] {receiver plane $40$~m};
% source
\coordinate (S) at (65,120);
\fill[red!70!black] (S) circle (2.2);
\node[above=3pt, red!70!black] at (S) {source, $z=120$~m, $L_W=80$~dB(A)};
% cardioid hint + thrust axis
\draw[red!70!black, ->] (S) -- ++(0,-16) node[below left=0pt] {\scriptsize thrust axis};
\draw[red!50] (S) ++(0,-4) ellipse (10 and 7);
% direct + diffracted path over a rooftop
\draw[->, thick] (S) -- (160,40);
\node at (150,66) {\scriptsize direct path};
\draw[->, thick, dash dot] (S) -- (112,70) -- (128,40);
\node at (98,86) {\scriptsize Maekawa / Fresnel $N_f$ per band};
% shadow zone
\draw[gray] (120,3) -- (138,3);
\node[gray] at (129,10) {\scriptsize shadow};
% wind + refracted ray
\draw[->, very thick, teal!70!black] (8,105) -- (34,105) node[midway, above] {\scriptsize wind $(u,v)$};
\draw[->, teal!70!black] (S) .. controls (30,86) and (16,66) .. (6,44);
\node[teal!70!black] at (26,58) {\scriptsize PE refraction ($\le 500$~Hz)};
% annotations
\node[align=left, anchor=north west] at (132,118) {\scriptsize spherical spreading\\[-2pt]\scriptsize ISO 9613-1 absorption};
\end{tikzpicture}
\caption{Geometry of the Task 5 reference acoustic model: a monopole
source $120$~m above the acoustic ground with cardioid directivity, a receiver
plane at $40$~m, per-band Maekawa diffraction over the LBM-LES building-height
profile, and a parabolic-equation wind-refraction term ($\le 500$~Hz) driven
by the simulated horizontal wind.}
\label{fig:task5_schematic}
\end{figure}

\begin{figure}[b]
    \centering
    \includegraphics[width=\linewidth]{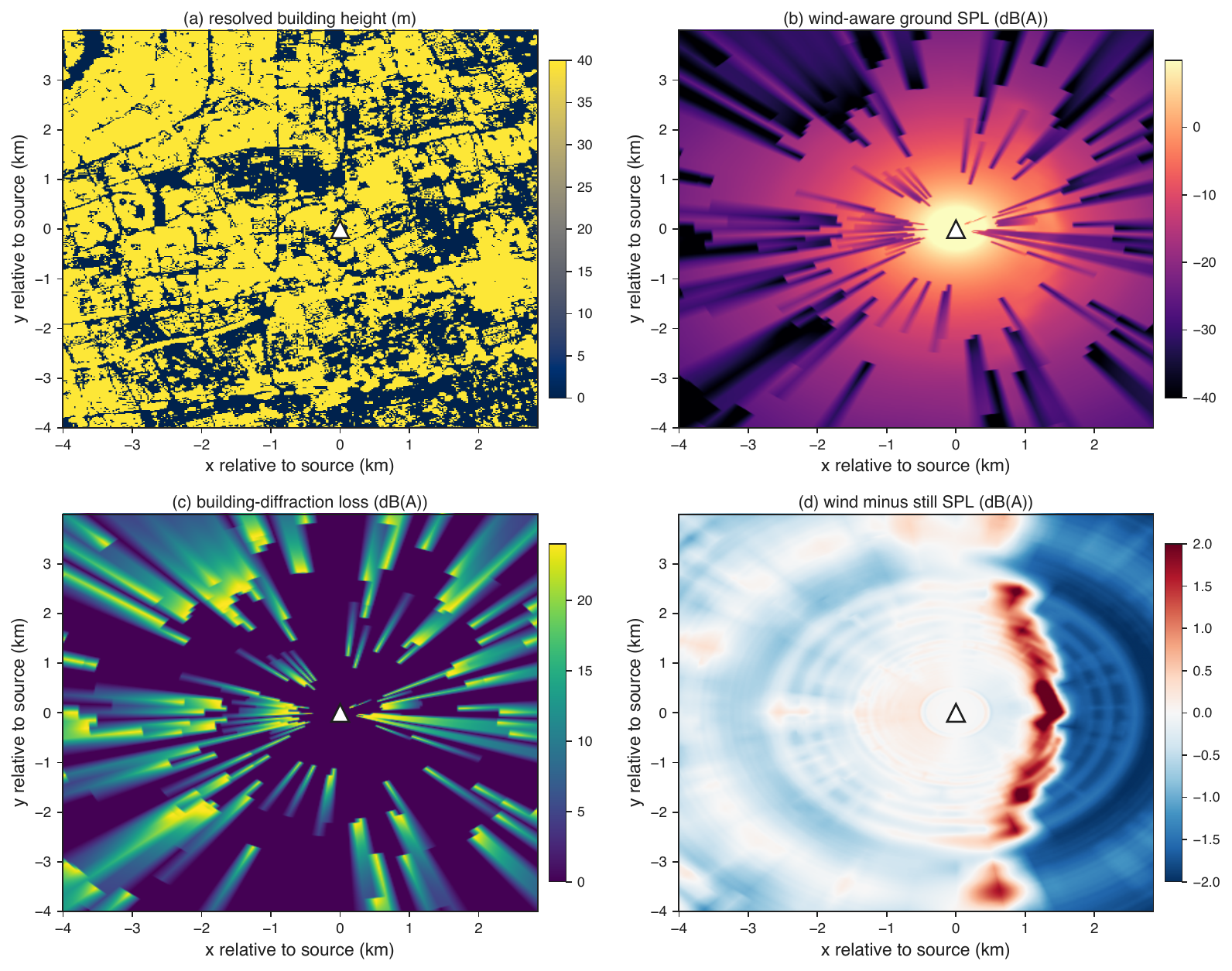}
    \caption{Task 5 UAV noise propagation for the easterly inflow case (ANG\_90) on the Shanghai \textsc{U3DWind} release. Panel a is the resolved building height above the terrain top, which is taken as the acoustic ground; the white triangle marks the source ground track. Panel b is the wind-aware ground SPL at the receiver plane ($40$\,m above the acoustic ground), from free-field spreading and ISO 9613 absorption minus the Maekawa building-diffraction loss plus the parabolic-equation wind term. Panel c is the building-diffraction attenuation in A-weighted dB(A), capped at $24$\,dB(A); panel d is the signed wind modulation $L_{p,A}^{\mathrm{wind}} - L_{p,A}^{\mathrm{still}}$ (color scale clipped at $\pm 2$\,dB(A)). The building diffraction is the dominant, deeply structured effect, while the wind adds a secondary upwind-shadow and downwind-enhancement asymmetry of a few dB(A).}
    \label{fig:task5_noise}
\end{figure}

An ISO 9613 model that accounts only for geometric spreading and atmospheric absorption, with no buildings, accumulates a mean error of $\varepsilon_{L_2}^{\mathrm{SPL}} = 4.645$ dB(A) and a maximum deviation of $26.130$ dB(A) against the reference, with a Shadow IoU of $0.000$ because it predicts no shadow at all. The still-air ray baseline, which replaces the graded Maekawa diffraction with a binary line-of-sight occlusion, barely improves on it, reaching $\varepsilon_{L_2}^{\mathrm{SPL}} = 4.564$ dB(A) with a Shadow IoU of $0.052$ and a deep-shadow IoU of exactly $0.000$. This is the designed failure mode of binary occlusion: a hard blocked or unblocked penalty cannot reproduce the graded, height-dependent diffraction depth, so it misses the deep shadow cores entirely. Both physics baselines ignore the wind and therefore share the constructional Wind RMSE of $4.040$ dB(A), which is the RMS magnitude of the reference wind term itself. The full reference solver, building diffraction plus wind, defines the metric origin.

Wind-blind neural baselines must learn the footprint from the building mask and geometry alone. The wind-blind MLP and 2D-UNet reach $\varepsilon_{L_2}^{\mathrm{SPL}}$ of $4.007$ and $4.130$ dB(A) but Shadow IoUs of only $0.032$ and $0.000$: they fit the smooth range-dependent structure of the field, yet essentially fail to localize the diffraction shadows, and their maximum deviations stay above $23$ dB(A). Predicting no wind modulation at all, they share the constructional Wind RMSE of $4.040$ dB(A) with the physics baselines that ignore the wind.

Providing the resolved wind field helps most where it should. The wind-aware 3D-UNet is in a near-tie with the wind-blind MLP on the total SPL error ($\varepsilon_{L_2}^{\mathrm{SPL}} = 3.968$ versus $4.007$ dB(A)), but its Wind RMSE of $3.211$ dB(A) sits clearly below the constructional $4.040$ dB(A) floor, showing that it extracts genuine refraction information from the flow field, and its Shadow IoU of $0.197$ begins to localize the shadows. The wind-aware FNO is the best pure-learned baseline on the total error ($\varepsilon_{L_2}^{\mathrm{SPL}} = 3.051$ dB(A)), with a Shadow IoU of $0.328$ and a deep-shadow IoU of $0.492$, because its spectral convolutions preserve the global shadow structure, although its Wind RMSE of $3.433$ dB(A) still leaves much of the refraction unmodelled. Recovering the deep diffraction shadow and the weaker wind modulation simultaneously from a feed-forward operator remains an open challenge and motivates the hybrid approach.

The diffraction-aware ray surrogate is the strongest baseline on every accuracy metric, with $\varepsilon_{L_2}^{\mathrm{SPL}} = 1.821$ dB(A), $\Delta L_{p,A}^{\max} = 8.337$ dB(A), Shadow IoU $0.397$, and deep-shadow IoU $0.749$. By solving the analytic still-air diffraction field inside its prediction path and learning only the wind-refraction residual, it anchors the building-shadow geometry to the physics; its SPL error therefore equals its Wind RMSE of $1.821$ dB(A) by construction, and that value is less than half the $4.040$ dB(A) floor of the wind-ignoring baselines. Hybridizing the diffraction physics with a learned wind residual remains the most reliable design on this task, although the residual wind error shows the refraction term is only partially captured.

Most importantly from an operational perspective, the surrogate resolves the cost bottleneck of building-aware acoustic simulation. Generating a reference label requires $487.6$ s per case, of which $484.1$ s is the Crank-Nicolson parabolic-equation wind march while the still-air diffraction split alone takes $3.5$ s, a cost that is computationally prohibitive for dynamic, city-wide trajectory optimization. In contrast, the diffraction-aware ray surrogate produces a wind-aware footprint in $3.145$ s, an acceleration by a factor of $155.0$, while retaining the best accuracy of all baselines; the pure-learned baselines are faster still, at $0.224$ to $1.050$ s per case (factors of $464.5$ to $2177.6$), but at the cost of losing the shadow geometry. This efficiency makes it feasible for UAM traffic management systems to re-evaluate and update community noise footprints in near real time as local wind conditions shift.

Fig.~\ref{fig:task5_noise} shows how the building canopy and the wind shape the footprint for one source on the easterly inflow case. The resolved building heights (Fig.~\ref{fig:task5_noise}a) drive the wind-aware ground SPL at the receiver plane (Fig.~\ref{fig:task5_noise}b): the field is far from a smooth radial pattern, carved into bright corridors and deep shadows by the canopy. Panel (c) isolates the building-diffraction attenuation, which reaches the $24$ dB(A) Maekawa cap directly behind the tallest towers and falls to zero in open sky, a graded structure that the reference's Maekawa diffraction term resolves, the binary-occlusion still-air ray misses (Shadow IoU $0.052$, deep-shadow IoU $0.000$), and the pure neural operators only approximate. Panel (d) is the secondary wind modulation $L_{p,A}^{\mathrm{wind}} - L_{p,A}^{\mathrm{still}}$: an upwind reduction and a downwind enhancement of a few dB(A) growing with range, far smaller than the building shadows but in the physically correct direction.

\section{Benchmark protocol details}
\label{app:protocol}

Every benchmark number in this paper, except the path-planner comparison of Table~\ref{tab:task3_planners} (single deterministic runs retained from the initial evaluation), is produced by
\texttt{scripts/run\_benchmark.py} in the released repository
(\url{https://github.com/ATMSaxon/U3DWind-Benchmark}), which stamps a metric
version (\texttt{2026-07-29-revision}) and a full provenance block (config
hash, index hash, split label, seed, timestamp) into every
\texttt{metrics.json}; the paper tables are regenerated from those files by
\texttt{scripts/build\_tables.py}. One YAML per (task, model) pair in
\texttt{configs/} records every hyperparameter; Table~\ref{tab:protocol}
summarizes the splits and the shared training protocol. Splits are
deterministic given (task, split, seed); validation draws use a per-seed random partition of the non-test cases (nominally 85/15; realized counts are stamped into the provenance block of every metrics file). Experiments ran on
8$\times$RTX~4090D (24~GB) and 2$\times$A800 (80~GB) GPUs; multi-seed rows use
seeds 0--2.

\begin{table}[htbp]
\centering
\caption{Benchmark splits and shared training protocol. Case counts refer to
simulation cases; sample counts to evaluation units. Full per-model
hyperparameters are the \texttt{configs/} YAMLs in the released repository.}
\label{tab:protocol}
\footnotesize
\setlength{\tabcolsep}{3pt}
\begin{tabular}{l l p{0.13\textwidth} p{0.115\textwidth} p{0.21\textwidth} p{0.155\textwidth}}
\toprule
Task & Split & Train / Val & Test & Evaluation unit & Optimizer family \\
\midrule
1 Surrogate & IE & 125 cases (85/15) & 18 cases & full 3D field ($384{\times}384{\times}32$ crop, 40~m) & Adam(W), lr $10^{-3}$--$10^{-4}$ \\
2 Reconstruction & IE & 251 cases (85/15) & 36 cases & field at $N_s \in \{16, 64, 256\}$ sensors & Adam(W), lr $10^{-3}$ \\
3 Risk & IE & remaining corpus (85/15) & 18{,}000 trajectories & per-trajectory risk score & Adam(W), lr $10^{-3}$ \\
4 Ranking & SO (seed 0) / CC & remaining slots / Shanghai & 64 / 144 queries & per-query candidate ranking & per-model (tree, MCDA, torch) \\
5 Noise & IE & 96 cases (85/15) & 47 cases $\times$ 16 sources & 752 SPL footprints & Adam(W), lr $10^{-3}$ \\
\bottomrule
\end{tabular}
\end{table}

\end{document}